# Decoding Segregation:
## Navigating a century of segregation research across disciplines and introducing a bottom-up ontology

Vini Netto, Kimon Krenz, Maria Fiszon, Otávio Peres, Desirée Rosalino

**Abstract**
Segregation is a widely recognised phenomenon with profound implications for societies worldwide. From political science and gender studies to anthropology and urban studies, it has garnered considerable attention across numerous scientific fields due to its multifaceted nature. However, what makes segregation such a far-reaching phenomenon? In fact, how many forms of segregation exist? Have different disciplines engaged in segregation research uncovered all its facets? This article systematically explores the landscape of segregation research spanning over a century. We analyzed 10,754 documents from the Scopus database to unveil the dynamics of the discovery of segregation forms through several findings. We identify (1) the exponential growth and increasing diversification of segregation forms, driven by combinatorial and exploratory work and increasing transdisciplinarity and intersectionality in research; (2) the evolution and structure of the field in hierarchies and clusters of segregation forms, revealing trends, persistence, and shifts over time; (3) the timing and geographical distribution of first publications on segregation forms, along with contextual variations across world regions and countries; (4) path dependencies in the historical and geographical shaping of segregation research; and (5) the structure of knowledge production. Aiming to contribute semantic organization to an increasingly complex field, we explore these findings to introduce a bottom-up ontology of segregation, marking the first comprehensive effort of its kind.

**Keywords:** segregation, segregation forms, segregation research, bottom-up ontology.

## Introduction

> Segregation is a multi-dimensional process, requiring a multidisciplinary approach.
> Vaughan and Arbaci (2011)

Segregation is a well-known phenomenon that probably impacts billions living in complex societies worldwide. Institutionalised in daily practices and the form of law as recently as in the early 20th century, it has gathered widespread public and academic attention. Segregation is a crucial subject across dozens of scientific fields, from Sociology and Urban Studies to Environmental Science and Gender Studies (Fig. 1a). What makes segregation such a far-reaching phenomenon, sweeping into so many aspects of social reality and different areas of knowledge? First of all, what is segregation? Definitions may vary. Segregation refers to processes, practices or situations in which individuals or social groups are separated, or their interactions restricted, based on distinguishing characteristics such as race, ethnicity, income, gender, occupation, education, geography, age, or other social attributes. Segregation can manifest in various degrees and forms, arising from voluntary choices, contextual conditions, or imposed restrictions. It is driven by a complex interplay of social, economic, political, and cultural forces. The nuances of segregation often involve both structural and relational dimensions, making it a multifaceted and dynamic phenomenon (see Freeman, 1978; Massey & Denton, 1989; Vaughan & Arbaci, 2011). The variability in how individuals and groups are classified, and the fluidity of such boundaries, add to the complexity of understanding segregation, which spans across multiple domains and scales. There is growing attention to the multidimensionality of segregation as a phenomenon that manifests over people and places, situations and contexts and through different means and materialities (Dadashpoor & Keshavarzi, 2024; Goldhaber & Schnell, 2007; Musterd, 2020). However, how does segregation manifest across social and material reality? How many forms of segregation exist? Have the sciences engaged with the problem brought all its facets and manifestations to light?

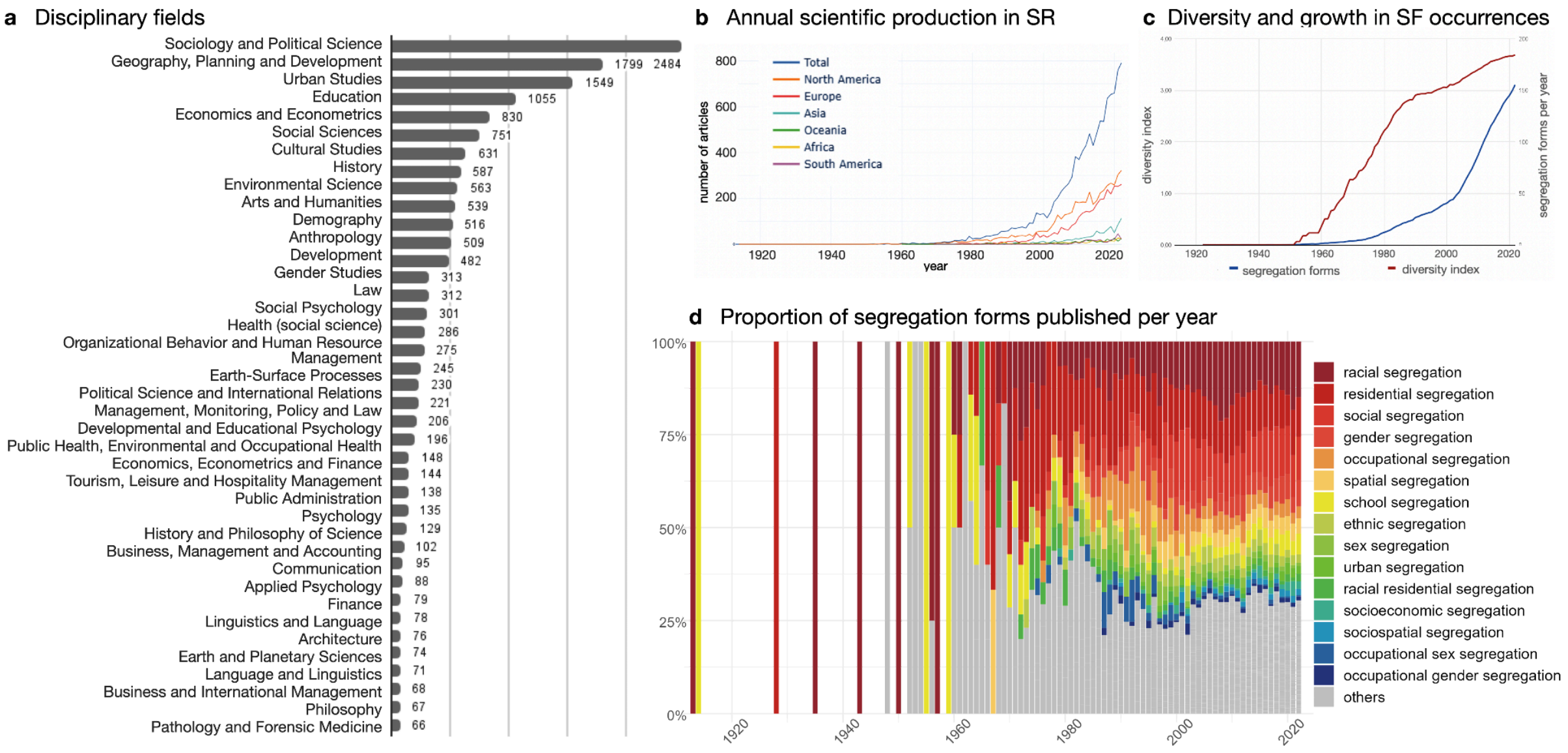

Fig. 1 – (a) The 40 disciplinary fields that most published articles on segregation in human affairs from a pool of 169 scientific areas available in the Scopus database (Table ST01 and graph G01 in the supplementary information[1]). (b) The number of publications per year and continent (1913-2022). (c) The growing diversity of segregation forms (SFs) published per year (red line, six years moving average) and the exponential growth in the number of SFs published per year (blue line, six years moving average). (d) Proportion of SFs published annually shows the top 15 forms with the highest publication numbers (G02).

Due to its pervasive nature, segregation is naturally hard to pinpoint. It is usually dealt with analytically in specific social and material manifestations. Its multidimensionality is apparent in titles, keywords and manuscripts of scientific papers and books as they address associated forms, means and expressions of segregation. Its complexity is also evident in the frequent conflation of terminology. Quite different terms are used interchangeably, such as "segregation", "urban segregation" and "socioeconomic segregation", frequently used to address residential segregation. "Neighbourhood segregation", "urban segregation", "territorial segregation" and "spatial segregation" may refer to the same segregation form. In turn, a single term may refer to different things – say, "spatial segregation" can be used to address physical barriers such as walls or distant locations within a city. Recent works have been addressing a frequent lack of precise definitions (e.g. Dadashpoor & Keshavarzi, 2024; Musterd, 2020). On the other hand, the multidisciplinarity involved in catching up with such complexity does not make the problem easier. Works in different disciplines address different segregation forms, sometimes using similar terminology. "Vertical segregation" has been used at least since 1986 to address the unequal distribution of demographic groups across different hierarchical positions within an organisation or industry (Corcoran, 1986), but also to address social differentiation in high-rise neighbourhoods and buildings (Maloutas, 2024; Maloutas & Karadimitriou, 2001). "Micro-segregation" is a growing subject in urban segregation research (e.g. Flint et al., 2012; Maloutas & Karadimitriou, 2022), even though the term initially appeared in the context of gender occupational segregation research (Jerby et al., 2006). In addition, growing attention has been paid to combinatorial and intersectional aspects of segregation over the past few decades. For instance, "occupational segregation" appeared in 1975 (Armstrong & Armstrong, 1975; Fottler, 1976). "Gender segregation" appeared in 1985 (Harkness & Super, 1985) and also in the work context (Brooker-Gross & Maraffa, 1985; Jones, 1985). Both terms intersected as "occupational gender segregation" in 1987 (Fox & Fox, 1987).

In short, segregation is a highly diverse phenomenon that manifests in different instances and through different means captured by different sciences. Studies within specific disciplinary fields are limited in their ability to address the possibilities of connections among the rapidly expanding array of segregation forms explored in an extensive – and still growing – body of literature (Fig. 1b–d). This empirical challenge is followed by epistemological difficulties and frequent terminological conflation in multidisciplinary literature. In this work, we introduce a means to systematically analyse how different manifestations of segregation have been covered across the social sciences, as available in the database of academic publications Scopus. Combining automated scientometric instruments and interpretive work, we map the landscape of segregation research (SR) and the trajectories of segregation forms (SFs) over a century across disciplines. We identify:

1. **The exponential growth and diversification of SFs** in the literature across 169 disciplinary fields.
2. **The dynamics of identification of segregation forms over time:** how new forms of segregation are discovered through combinatorial and exploratory scientific work.
3. **Relations of segregation forms and the structure of the field:** the connectivity between segregation forms in co-occurrence networks within the analysed literature.
4. **Multidisciplinary, transdisciplinarity and intersectionality in SR:** how specific SFs cut across disciplines and which fields have a higher diversity of SFs.
5. **Contextual variations:** the role of context in SR and the dominant segregation forms in different countries and world regions.
6. **Path dependence:** the time of SFs' first publications and their position in co-occurrence networks as traces of path dependencies in the historical and geographical shaping of SR.
7. **The production of knowledge:** the structure of knowledge production in co-citation and collaboration networks across countries in the transdisciplinary field of SR.

[1] Supplementary Information published at https://doi.org/10.5281/zenodo.13886005

The topology of segregation revealed by these methods will support a proposition:

8. **A bottom-up ontology of segregation** as a means to semantically organize access to this vast transdisciplinary landscape, deploying network analysis and AI-based semantic generation.

Of course, there are many ways to develop an ontology (Gruber, 2016; Hedden, 2016). We opt for an inductive, bottom-up approach instead of a top-down approach based on authors' contact with and interpretation of the scientific landscape (e.g. Dadashpoor & Keshavarzi, 2024). The bottom-up approach can reduce interpretive noise while increasing consistency in scrutinising concepts both panoramically and in-depth, particularly considering a strongly multidisciplinary field that could hardly be fully accessible for individual researchers. We devised a method to exhaustively search for terms associated with segregation in the Scopus database and semantically identify the segregation forms addressed by authors. The works of decoding segregation through mapping the research landscape and proposing an ontology ordering segregation forms based on relationships and meanings explored in the literature are coupled with introducing an online collaborative platform. Segregation Wiki brings the 804 segregation forms identified by our meta-study and adds descriptions. We shall now briefly revisit previous efforts to identify the multiple dimensions of segregation.

# Reviews and classifications in segregation research

There is growing attention to the diverse nature of segregation as authors try to make sense of its manifestations and multidimensionality. Many works have focused on providing a global perspective on segregation forms, concepts and measures. Massey and Denton (1988) introduced dimensions of segregation as problems of measures of evenness, exposure, clustering, centralisation and concentration. Some works explore data-based segregation indices related to different variables, including residential segregation involving ethnicity, religion, income, and age as segregation features (e.g. Cottrell et al., 2017). Many segregation forms, like social segregation, spatial segregation, intersections of class, ethnicity and age (Boterman, 2013), socioeconomic, income, class, education and race, intergenerational segregation, and individual exposure, were addressed in Musterd (2020).

Indices of urban segregation were also organised according to temporal evolution in generations of measures: a first generation based on non-spatial representations, a second based on spatial analyses, and a third focused on people, individual behavior and time (Feitosa et al., 2023; Kwan, 2009, 2013). Segregation was analysed as a multidimensional characteristic of individual behavior based on different types of segregation: ethnic, residential, territorial and interactive segregation in daily activity spaces (Goldhaber & Schnell, 2007) and social networks, first addressed through topological centrality measures by Freeman (1978). Gauvin and Randon-Furlin (forthcoming) address several types, definitions and dimensions of segregation – namely, social, spatial, residential, and dynamic segregation. They identify a set of segregation indices such as network measures in social segregation and dissimilarity, multigroup, and multiscale measures for estimating spatial segregation. Such reviews are generally based on top-down scannings of existing literature, meaning that authors search for segregation forms they pre-identified throughout their interpretive contact with the literature over time. Systematic overviews stem mainly from methodological works on segregation measures.

Recent efforts have been made to map the literature bottom-up through broad scanning. Müürisepp et al. (2022) focused on segregation in activity spaces, and mobility flows to analyse ethnic, racial, religious, linguistic, socioeconomic and demographic dimensions of segregation. Dadashpoor and Keshavarzi (2024) define urban segregation concerning other pre-selected forms like residential segregation, spatial segregation, activity space segregation and workplace segregation defined in a top-down manner over 182 articles selected over an initial sample of 3,595 articles in the Scopus Database. As laudable efforts, these works still seem narrow in their disciplinary coverage of the enormous possibilities of segregation forms. Purely interpretive top-down efforts to identify the plethora of segregation forms can hardly cope with the diversity of forms already explored across dozens of scientific fields. The field of segregation research would benefit from a more encompassing and transdisciplinary mapping of the multiple segregation forms, i.e. a conscious effort towards the enumeration and classification of segregation forms identified in over 110 years of scientific production. A way to do this is through a *bottom-up ontology* considering segregation forms identified in the broadest possible spectrum of disciplinary fields.

# Materials and methods

We explored the landscape of segregation research through a scientometric method, digitally analysing an extensive database of articles, chapters and books. The Scopus database covers almost 18,000 titles from over 5,000 publishers. We cannot directly infer the quality and representativeness of the database as a sample of the totality of segregation research (see Supplementary Information - SI). However, the considerable overlapping between digital repositories Scopus, Google Scholar and Web of Science, with high Spearman correlations between citation counts (0.78-0.99) (Martín-Martín et al., 2018), suggests extensive coverage. Scopus is reliable as a database, encapsulating a substantial part of published scientific works. The database used in this work contains all variables available in Scopus except the core manuscript text, including: “Authors”, “Titles”, “Year”, “Source.title", “Volume”, “Issue”, “DOI", “Cited.by", “Affiliations", “Abstract", “Indexed.Keywords", “Funding.Details", “References", “Correspondence.Address", “Editors", "Publisher", "Conference.name", "Conference.location", "Conference.date", "ISSN", "ISBN", "CODEN", "Language.of.Original. Document", "Document.Type" and “EID”. We selected only documents containing the word "segregation" at least in one of the fields "Title", "Abstract" and "Keywords" from different scientific areas: Social Sciences, Arts and Humanities, Economics, Econometrics and Finance and Psychology (for the complete list of disciplinary fields, see Table ST01 in SI). Publications from journals in the Life, Health, and Physical Sciences were excluded from the analysis, as the term "segregation" in these fields – such as in genetic segregation, acoustic signal segregation, or segregation in neuronal activity – typically refers to biological or technical processes, rather than societal or social phenomena. Eventually, journals and works within those sciences were considered as they are assigned multi-disciplinary fields, including in the Social Sciences. Our resulting dataset contains 3,463 different sources from 1913 to 2022. We selected 10,754 documents containing the word "segregation", which include 8,483 articles, 1,196 book chapters, 322 books, 124 conference papers, 36 editorials, and 593 reviews of original research.[2] All works are written in English for both feasibility and international reach. We are aware of the significant limitation that such a linguistic scope imposes (see Acknowledgements). As we used titles, keywords and abstracts as our information sources, we included books, book chapters and conference papers offered by Scopus with that information available. The selected documents feature 14,841 authors and include 473,706 references spanning works inside and outside the corpus.

## Identifying segregation forms in the literature

The method for identifying segregation forms represents them as *n-grams*, a sequence of *n* symbols or units of text in a particular order found in a language dataset. A unigram contains one term (e.g. "segregation"), a bigram contains two terms (e.g. "occupational segregation") and so forth. Our method involved (a) an automated procedure to select n-grams of potential interest, (b) an interpretive procedure to assess whether n-grams can be considered segregation forms, and (c) quantitative and semantic analyses of the validated segregation forms to identify ontological relationships (Fig.2).

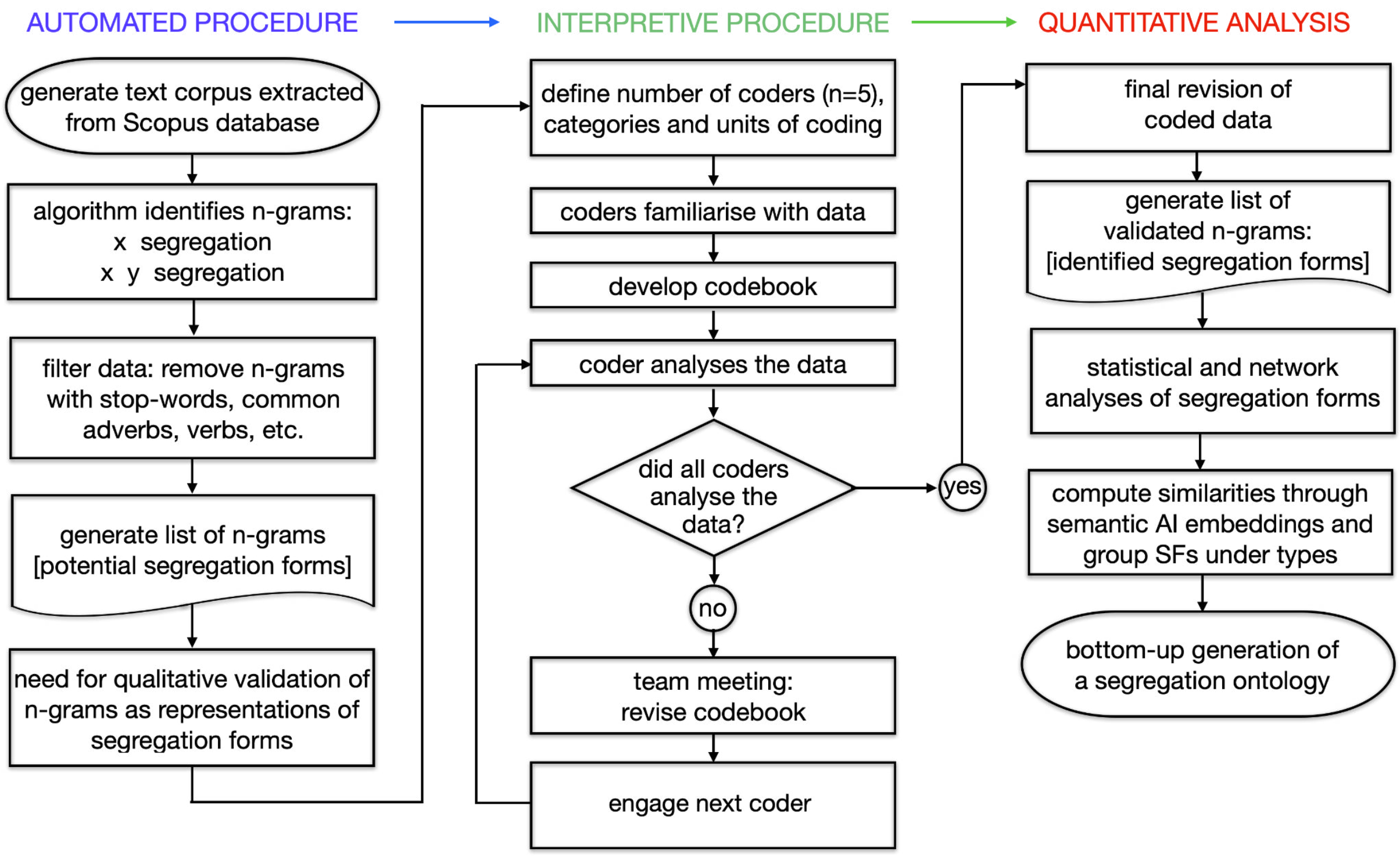

Fig. 2 – Method: from automated to interpretive procedures to support the analysis of the multidisciplinary field of segregation research and the inductive generation of an ontology of segregation forms.

### *(a) Automated procedure*

We analysed the text corpus derived from the Scopus database using a Python algorithm coded to search for terms associated with "segregation" considering the syntax of sentences in English, namely including one or two preceding words:

x segregation (bigram)

[2] According to the Scopus definition, a review consists of a "Significant review of original research, also includes conference papers. Characteristics: Reviews typically have an extensive bibliography. Educational items that review specific issues within the literature are also considered to be reviews. As non-original articles, reviews lack the most typical sections of original articles such as materials & methods and results." See page 11 of the Scopus content guide at https://www.elsevier.com/products/scopus/content.

x y segregation (trigram)

"x" and "y" can capture words within texts that might qualify segregation forms (e.g. "social segregation") but also terms unrelated to meaningful segregation forms (e.g. "whether segregation"). We are aware that such construction may limit the identification of segregation-related n-grams that do not use such a usual syntactic construction (e.g. "segregation in urban trajectories") but estimated that their occurrence is relatively rare compared to the more common structure outlined above. Trigrams potentially combine two bigrams (i.e. "x y segregation" combines "x segregation" and "y segregation", e.g. "occupational gender segregation" combines "occupational segregation" and "gender segregation"). That fact required attention when we assessed frequencies of n-grams. Firstly, to avoid duplicating SFs in sets of trigrams and bigrams, we applied an automated procedure to remove terms identified as part of trigrams from the set of bigrams, e.g. once the term "occupational gender segregation" was computed as a trigram, the words "gender segregation" it contained were no longer computed as a bigram for that particular occurrence. Secondly, concerning the *order* in which terms should be identified and counted (e.g. "vertical residential segregation" and "residential vertical segregation"), we considered both n-grams as equivalent and prioritised the one published earlier (i.e. based on the first year of publication). In the case of finding reversed trigrams first published in the same year, we considered the number of appearances of the term in the sample (i.e. a larger number indicates a more used term). Fourgrams were omitted from the analysis because the potential to add SFs was anticipated to be minimal compared to the interpretive effort required to discern SFs from all four-word phrases ending in "segregation".

The list of bigrams and trigrams identified by our algorithm within the corpus contained substantial noise. We removed bigrams and trigrams whose initial terms had no lexical meaning and those that primarily express grammatical relationships between words, function words and stop-words, e.g. "I", "you", "to", conjunctive adverbs (e.g. "accordingly", "furthermore", "moreover"), subordinating conjunctions (e.g. "after", "although"), auxiliary verbs (e.g. "could", "should", "would"), most common verbs (e.g. "do", "know", "like"), most common adverbs (e.g. "accidentally", "actually", "afterwards"), and scientific jargon (e.g. "et al.", "proceeding", "fig"). This filter resulted in a list of 5,635 n-grams (1,748 bigrams and 3,617 trigrams) containing "segregation" as their last word and potentially referring to segregation forms.

### *(b) Interpretive procedure: coding n-grams*

The remaining 5,635 n-grams still had to be examined regarding the conceptual sense of "x" and "x y segregation". To our knowledge, no automated procedure is available to sufficiently accurately disclose the meaning of the remaining n-grams to identify which refer to empirically relevant segregation forms. Therefore, this step involved interpretive access to every remaining n-gram as a potential segregation form. This task presents a significant challenge, primarily because of the sheer size of the subset under consideration. Additionally, potentially valid n-grams were not always explicitly categorised as forms of segregation in the original works from a conceptual standpoint. We do not expect authors to declare segregation forms conceptually as they can meaningfully deal with them empirically or theoretically without explicit definitions. This issue also poses the interpretive problem of the subtle line between forms of segregation meaningfully dealt with and sheer incidental terms. A strong example of the first case is the term "hypersegregation" (Massey & Denton, 1989), explicitly proposed as a form of segregation – as opposed to the incidental case where terms might appear only once in an abstract or as a keyword but are neither present in the main text nor explored as segregation forms. We further checked terms that fell in the second condition by assessing the literature outside Scopus, namely covered by Google Scholar as a second database, and validated them in the case of meaningful use.

Of course, deriving textual meanings from a given text segment requires a certain level of interpretation (Krippendorff, 2004). A systematic analysis requires a rigorous interpretive coding framework diligently applied to qualitative data (M. W. Bauer, 2000; Hruschka et al., 2004). One way of reaching consensual interpretations is by having different researchers participate in coding work sequences. This approach ensures that the underlying analytical structure has a meaning that extends beyond an individual researcher's perspective (O'Connor & Joffe, 2020). Accordingly, procedure 2 involved a team-based approach to interpreting the variety of terms algorithmically captured. We opted for an iterative thematic analysis (O'Reilly, 2012) to achieve inter-coder transparency, consistency and reliability (Hemmler et al., 2022). The sequential procedure allows the repetitive evaluation of the quality of the coding process, building coherence and stability of interpretations across the research team (Cascio et al., 2019).

The iterative coding process involved five researchers active in segregation research (Fig.2). Initially, coders familiarised themselves with the varied types of qualitative data. Then, they developed a codebook as a coding frame with criteria and guidelines for individual interpretation. Codes were used as discrete categories to analyse n-grams as representations of segregation forms according to three states: a valid n-gram, a non-valid n-gram or an n-gram requiring discussion. Coders used online sheets with shared access to code the qualitative data. To verify whether n-grams containing the word segregation as the subject referred to segregation forms, they analysed the articles' abstracts, titles and keywords. When coders deemed these pieces of information insufficiently informative, they accessed the respective manuscripts. They also added comments to the coding sheet for collective discussion. After every individual coding round, coders had a full view of previous coding work, allowing them to learn about doubts and observations of the prior coder(s). Team meetings followed each iteration of coding work to check whether decisions were deemed appropriate and to identify items of discrepancy. Each coder could point out discrepancies by reviewing previous coders' coding choices. Differences were discussed to reach collectively accepted decisions. Such decisions were used to iteratively update the codebook and then apply it in the following coding round (Fig. 2) (see the Codebook[3]).

Coders validated 804 n-grams (371 bigrams and 433 trigrams) (see Table ST03). Nearly half of the validated n-grams (412 or 51.2 % of the sample) appeared only in one document within the sample, whereas 15.2% appeared in two documents. We chose to keep n-grams with only one publication in the sample instead of opting for a higher frequency threshold due to our interest in grasping potentially new segregation forms as they are published and become part of the literature.

### *(c) Quantitative analyses*

We deployed a set of statistical methods and network analyses, which will be explored in the next section.

# Analysis and results

## 1. The **exponential** growth and diversification of segregation forms

How has research on segregation evolved? What were the first segregation forms (SFs) identified, and where were they identified? Publications on forms of segregation such as racial segregation, school segregation and residential segregation can be found in non-academic literature at least as early as the 17th century.[4] A particular peak in publication coincides with the mention of these three early identified SFs in the Records and Briefs of the United States Supreme Court (1832). Documents of investigative nature addressing segregation can be found by the end of the 19th century, such as Charles Booth's poverty maps in London (1889) and Emily Balch's (1895) review of the Hull Maps and

[3] Codebook for identifying segregation forms published at https://doi.org/10.5281/zenodo.13886304
[4] See Google Books Ngram Viewer: http://bit.ly/3xFfH9k

Papers, "a presentation of nationalities and wages" in a district of Chicago.[5] We shall focus on academic publications available in Scopus. The database includes documents from 1913 onwards (table 1). To be sure, publications outside the database might precede the first publication of segregation forms identified in Scopus. However, we adopted the first year and country of publication in papers for methodological consistency as SFs appeared in the Scopus database (see Acknowledgements). For instance, the earliest academic works identified to deal with racial segregation, school segregation and residential segregation are respectively from 1913 (Otis, 1913), 1914 (Gault, 1914) and 1928 (Burgess, 1928) (Table 1).

| | Segregation Form | Country of first publication | First year of publication | Number of publications |
|---|---|---|---|---|
| 1 | racial segregation | United States | 1913 | 2075 |
| 2 | school segregation | United States | 1914 | 558 |
| 3 | residential segregation | United States | 1928 | 1810 |
| 4 | enforced segregation | United States | 1948 | 4 |
| 5 | physical segregation | United States | 1952 | 30 |
| 6 | institutional segregation | South Africa | 1953 | 27 |
| 7 | compulsory racial segregation | United States | 1954 | 1 |
| 8 | ecological segregation | United States | 1954 | 6 |
| 9 | group segregation | United States | 1954 | 25 |
| 10 | educational segregation | United States | 1956 | 89 |
| 11 | slow learner segregation | Australia, United States | 1960 | 3 |
| 12 | large scale segregation | United States | 1962 | 6 |
| 13 | chicago school segregation | United States | 1963 | 1 |
| 14 | de facto segregation | United States | 1963 | 78 |
| 15 | formal job segregation | United States | 1963 | 1 |
| 16 | ethnic congregation segregation | United States | 1964 | 1 |
| 17 | class segregation | United States | 1965 | 118 |
| 18 | housing segregation | United States | 1965 | 115 |
| 19 | racial residential segregation | United States | 1965 | 249 |
| 20 | neighborhood segregation | United States | 1966 | 86 |
| 21 | social segregation | United States | 1966 | 1708 |
| 22 | territorial segregation | United Kingdom | 1966 | 23 |
| 23 | spatial segregation | Australia | 1967 | 589 |
| 24 | negro residential segregation | United States | 1968 | 2 |
| 25 | urban school segregation | United States | 1968 | 3 |
| 26 | trolley car segregation | United States | 1969 | 1 |
| 27 | optional segregation | United States | 1971 | 1 |
| 28 | role segregation | United States | 1971 | 38 |
| 29 | sexual segregation | United States | 1971 | 38 |
| 30 | economic segregation | United States | 1972 | 154 |

Table 1 – The earliest 30 segregation forms identified in the Scopus database, the year and country of first publication, and the number of documents in the sample (see Table ST03).

Our sample suggests that, since its inception in academic research, the number of SFs examined in the literature has grown exponentially (Fig. 3a), with an average annual growth rate of 3.26% (Fig. 3b). The distribution of frequencies of publications is uneven across different segregation forms. As we shall see, different factors may explain such growth, mainly concentrated on some SFs – many of them historically under attention in the academic literature (Fig. 3d). Racial segregation, residential segregation, and social segregation appear in the largest number of papers. More recent SFs related to particular social realms, like gender, occupation, ethnicity and sex, and material forms (residential, spatial and urban segregation), are among the most frequent. Key socioeconomic features such as class and income appear from the 13th to the 21st positions. Bigrams make 31 of the 40 most published SFs. The most frequent trigram (racial residential segregation) is the 12th most frequent n-gram. The growth in the number of segregation forms identified by authors across disciplines naturally relates to the *diversification of the field*.

[5] We thank Laura Vaughan for bringing the Booth maps and Balch's article to attention.

We assessed how diverse the field has become over time using Shannon's (1948) entropy:

$$H_t = -\sum_{p_i=1} p_i \cdot ln(p_i)$$

Where $H_t$ is the diversity index of SFs per year, *In* is the natural logarithm, and $p_i$ is the proportion of publications of each SF in relation to the total number of SF published per year.

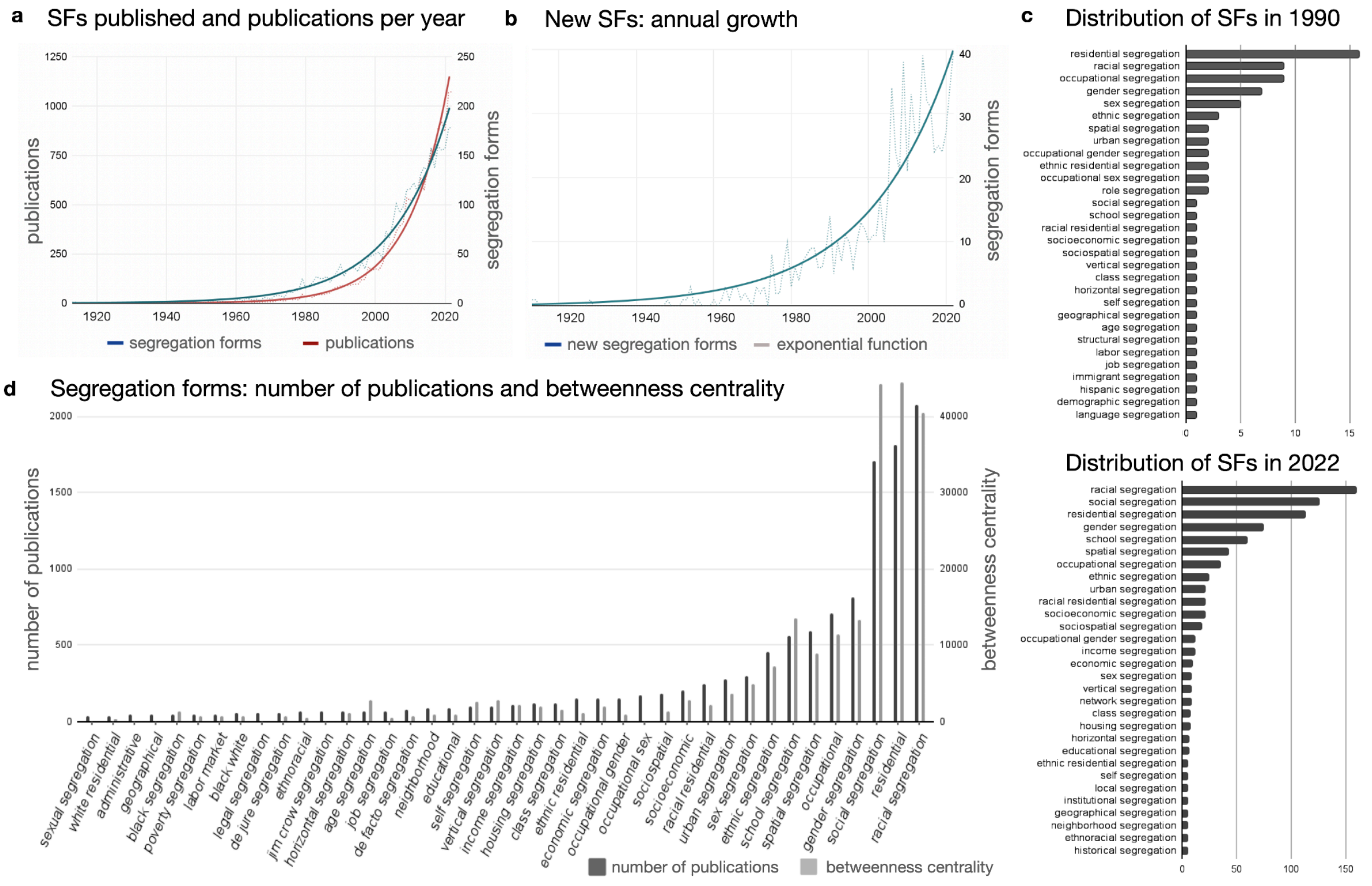

Fig. 3 – (a) Number of SFs published per year, with the discontinuous blue line showing the actual number of SFs and the continuous blue line representing the best fitting curve (y= (1.19E-51)e^0.0606x, R² = 0.979), and the number of publications in SR per year [red line, y = (1.21E-72)e^0.0854x, R² = 0.986]. (b) Number of new SFs introduced per year (dotted line, annual growth rate 4.83%) and the fitting exponential function (blue line, y = (4.66E-39)e^0.455x, R² = 0.897). (c) The distribution of SFs in 1990 and 2022 shows an increasing concentration around the most frequent SFs, which leads to a slower growth rate in the diversity of SFs (see Fig.1c). (d) Distribution of the total number of publications across segregation forms (top 30) in the sampled literature in 2022 (black bars) and betweenness centrality values (grey bars – see section 3).

The resulting curve (see Fig. 1c in the Introduction section, red line) shows strong growth in diversity, particularly from the 1950s onwards, slightly slowing down from the 1990s onwards. This is the case even though the total number of SFs per year continued to grow exponentially (Fig. 1c, blue line), along with the number of new SFs published per year (Fig. 3a) and the annual growth rate (Fig. 3b). The explanation for such apparently contradictory behaviors lies in the increasing concentration of papers published around certain SFs from the mid-1980s to the 2020s (Fig. 3c). We selected Shannon entropy as our measure of diversity due to its mathematical rigor in accounting for the distribution of events across different categories. Unlike simple measures of variety, Shannon entropy provides a probabilistic approach to diversity, capturing both the number of categories and the evenness of their distribution. This makes it especially useful for assessing the balance of publications across SFs. If one or a few SFs attract a disproportionately high number of publications, the overall growth in diversity may be constrained if compared to a scenario of equal distribution of new publications across SFs – even if the total number of SFs and publications continues to increase exponentially. Shannon entropy effectively captures these subtleties in the distribution dynamics. Summing up, new segregation forms continue to be identified. The field is becoming more diverse, even though diversification has slightly slowed down. But how does diversification happen?

## 2. The dynamics of identification of segregation forms

How can we explain the exponential growth and diversification of SFs over time? In fact, how have new forms of segregation been discovered? These questions go beyond the mere description of patterns of growth and diversification. They address the heart of production in science: the creation of ideas in the process of identification and conceptualisation of phenomena over time – in our case, new concepts able to grasp ever subtler forms of segregation. Of course, the genesis of concepts in science has been a research field in itself, at least since Beveridge (1957) and Merton (1957, 1961). Among the concepts developed to understand such dynamics, we explore Stuart Kauffman's (2000) evolutionary biology concept of "the adjacent possible". Once considered in the epistemological realm (Johnson, 2011; Monechi et al., 2017; Tria et al., 2014), the concept deals with the potential surrounding every idea: what is actual now enables what is next possible (Kauffman, 2000), a space of possibilities conditional to the occurrence of innovations. In essence, introducing a new idea creates a landscape of adjacent possibilities where even more innovations become thinkable (Monechi et al., 2017). An illustration of this is the fact that both Charles Darwin and Alfred Wallace independently developed the theory of natural selection after reading Malthus's *Principles of Population* (Beveridge, 1957). Darwin and Wallace's parallel insights are also a classic example of concurrent innovations that independent creators arrive at simultaneously, described by Merton (1961) as "multiple discoveries". The field of segregation research displays similar phenomena, as we shall see. While other mechanisms may also play a role, we hypothesise that the diversification of concepts have primarily two related processes as drivers (Fig. 4):

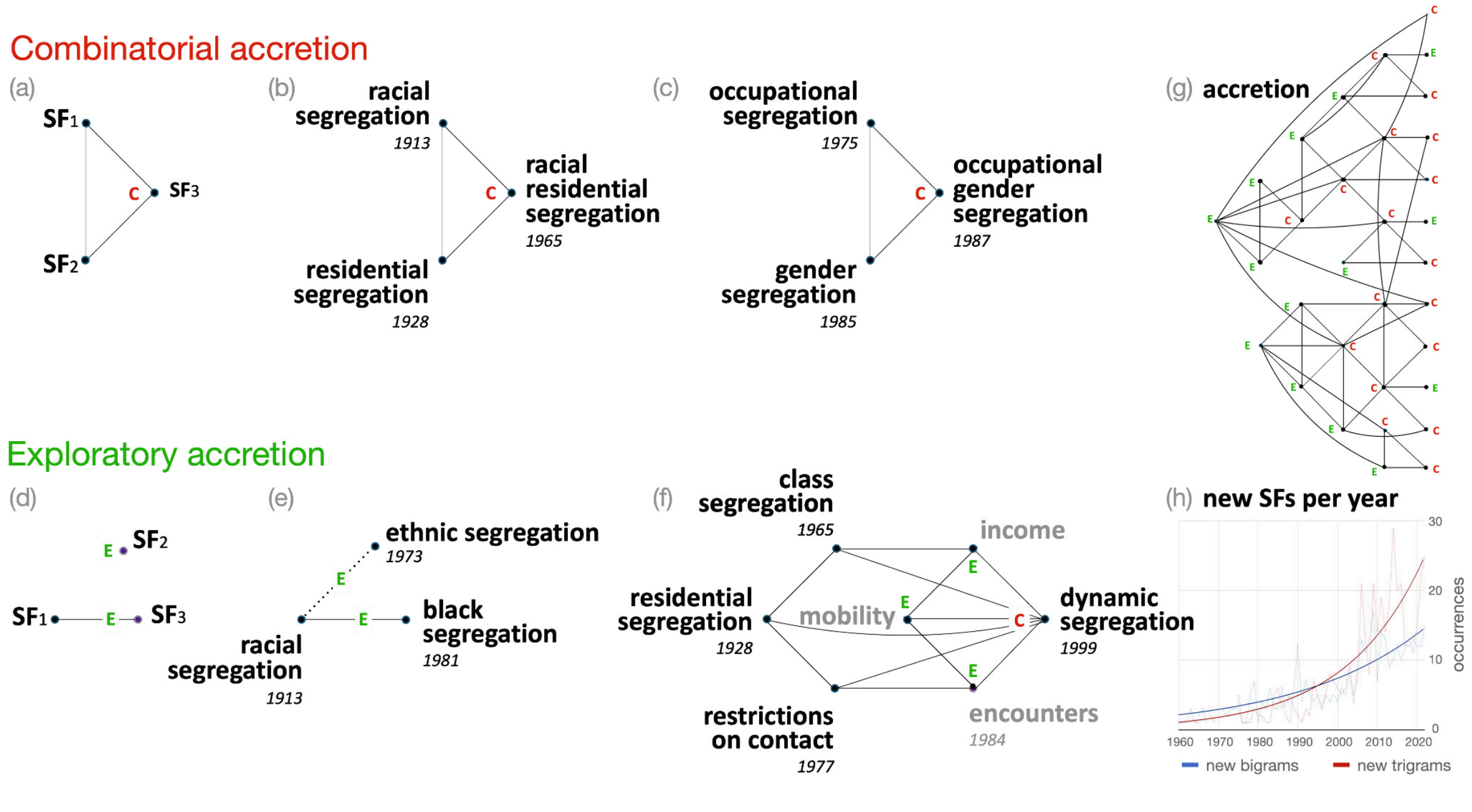

Fig. 4 – Conceptual graphs illustrating genealogies in segregation research (SR):
(a), (b), (c) represent combinatorial accretion, where known segregation forms are connected to form new combinations;
(d), (e) highlight exploratory accretion, which extends analysis into deeper dimensions of segregation;
(f) provides an example of how these two processes contribute to the identification of a new segregation form (SF);
(g) suggests how both accretion processes drive the exponential growth of SF numbers, including long-range connections over time. (h) Empirical data showing the increasing emergence of bigrams and trigrams representing new SFs over time, as hypothezised in (g).

(i) **Combinatorial accretion** achieves conceptual innovation through connections or analogies between two or more known objects or concepts (Beveridge, 1957). This process involves making new combinations of existing ideas or concepts, often drawing on previously unrelated domains. The capacity to proliferate variations, associations, or transformations from different elements – even from domains far apart (Boden, 2004) – leads to the emergence of novel concepts. This would be the case irrespective of the combinatorial procedure (from random, unconscious and undirected processes to systematically oriented associations) or the heuristics deployed (from blind ideational variations to analogy and trial-and-error) (Simonton, 2010). In segregation research, combinatorial procedures materialise through connecting known aspects or forms of segregation to identify new forms that manifesting through previously unrecognized means and situations or over previously unrecognized social realms or groups (Fig. 4a, b, c). In short, an identified segregation form brings latent or potential connections to other forms – connections which will eventually allow the discovery of a different, perhaps more specific, segregation form.

For instance, racial segregation and residential segregation have been among the first segregation forms explored in the scientific literature. The Scopus database includes 1913 and 1928 as their respective first publication years, the latter published by the Chicago School pioneer E.W. Burgess. This indicates that early segregation research often conflated multiple forms of segregation, as residential segregation inherently involved racial divisions, particularly in the American context. Growing social and political concerns with racial segregation associated with a judicial system and housing policies in the US in the following decades were explicitly addressed as "racial residential segregation" in the 1960s. The bigrams "racial segregation" and "residential segregation" appear, for instance, in Miller (1965), followed by an explosion of articles using the new trigram. Indeed, newly created combinatorial trigrams are particularly notable in their timing. Of the 224 trigrams formed from words also found in bigrams within the Scopus dataset, 204 (91.1%) were published after the bigrams, indicating that combinatorial accretion is a sequential and ongoing process in the identification of SFs (see Table ST04 in SI). The increasing frequency of trigrams (Fig. 4h) suggests that over time, analytical specification intensifies as new forms and subtler dimensions of segregation are identified, broken down or combined, and named.

(ii) **Exploratory accretion** involves the incursion in conceptual spaces or sets of possible ideas allowed by specific modes of thought (Boden, 2004). It involves both actively searching for previously unseen aspects of a phenomenon or entirely new phenomena, and serendipitous or unexpected encounters with ideas allowed by thought processes and knowledge. In the exploratory conceptualization of a phenomenon and its aspects, unrelated aspects identified elsewhere, like other research fields or disciplines, may trigger associations. Again, Darwin's and Wallace's use of Malthus's theory of population growth to imagine random biological changes active in natural selection and the emergence of new species is a classic example of exploring the adjacent possible through seemingly unrelated subjects. In segregation research, exploratory accretion may lead to identifying previously ignored social groups, experiences, material forms or scales – say, social networks segregated in digital media or the segregation felt at the individual's level. This process allows attention to ever more subtle forms of segregation.

For instance, the attention to racial segregation, at least since the 1910s in the US academic context, precedes the recognition of the nuanced diversity of ethnic segregation in the 1970s onwards (De Lepervanche, 1980) (Fig. 4e). Approaches to the segregation experienced by individuals in the here-and-now emerged in the late 1990s. A combination of residential and class segregation, limits imposed by income over people's mobilities, Freeman's (1978) view of segregation as "restriction on contacts", and a growing sensitivity to encounter as a feature of social reproduction (Giddens, 1984; Hillier & Hanson, 1984) led to the concept of "dynamic segregation" manifested as poorly overlapped networks of movement in public spaces structured by daily activities (Netto & Krafta, 1999, 2001; Netto et al., 2015; Fig. 4f). In the same year, analogous concerns with people's lifestyles and routines triggered the concept of "activity-space segregation" beyond residential spaces (Schnell & Yoav, 1999, 2001; cf. Kwan, 2009). A plethora of terms addressing these segregation forms and their factors have since emerged: individual-level segregation (Fineman, 2020; Feitosa et al., 2023), experienced segregation (Athey et al., 2020), spatiotemporal segregation (Liu et al., 2021), mobility segregation (Park et al., 2021), trajectory-based segregation (Malmberg & Andersson, 2023), etc. Exploratory accretion also includes situations where the identification of an SF does not explicitly involve an existing SF, like bigrams or non-combinatorial trigrams, e.g., attendance zone segregation, human capital segregation, nursing home segregation illustrate such process.

While these two processes are generally identified in the concept creation literature, we should allow for the possibility of other accretion processes. Indeed, more and more forms of segregation continue to be recognized in subtle realms and scales. The more SFs and knowledge, the more materials to combine and explore, triggering exponentiation as each new SF creates the conditions for the identification of new ones (Fig. 4g). Accretion by combination and exploration seems to be at the heart of processes of discovering new segregation forms — evolving ideas into the "adjacent possible". Combined, these processes explain the exponential growth and diversification of segregation forms as the field moves in multiple directions. However, what actual ramifications are produced from previous SFs or entirely new entries? How can we clarify the evolution of the field in its totality? More robust methods may throw light on the connectivity of segregation forms.

## 3. The network of segregation forms and the structure of the field

One way to explore the transdisciplinary topology of segregation forms (SFs) is through the examination of SFs mentioned together within the same work, referred to as "co-occurrences". Co-occurrence analysis, established in scientometrics since the 1980s (Bauin, 1986; Callon et al., 1991), is a method used to map scientific fields and identify patterns within the literature. It provides valuable insights into conceptual relationships, the genealogy of ideas, and the structure and dynamics of a research field (He, 1999; Sedighi, 2016). Co-occurrences (COs) reflect the frequency with which two or more terms (e.g., "socioeconomic segregation" and "spatial segregation") appear together in a specific text, such as an abstract, article, or set of documents. Authors may employ multiple segregation forms in a single work for various reasons: to explore their similarities, differences, or complementarities, or to highlight contrasts and oppositions. Alternatively, the use of different SFs could stem from conflation or misinterpretation, where distinct terms are used to refer to the same phenomenon. Regardless of the authors' intent, co-occurrences offer significant epistemological value. Even unintentional co-occurrences hint at potential associations, creating implicit relationships between terms. By doing so, they suggest that authors are, consciously or otherwise, engaging with different facets of segregation and their interconnections, offering readers new possibilities for understanding the topic.

The following co-occurrence analysis does not aim to explore the semantic connections between segregation forms (SFs) or the conceptual intentions behind their use, which would be better suited for qualitative or interpretive approaches. Instead, our focus is on exploring these connections to empirically assess the structure of segregation research and to better understand the process of identifying SFs within conceptual work. A co-occurrence (CO) network is a virtual construct, formed from pairs of SFs that appear together in individual works. Over time, these sets establish connectivity between documents that collectively shape the literature or field. By analyzing the literature as a series of overlapping documents, we can reconstruct these connections into a CO network. As SFs co-occur within works, their interrelations within the broader literature can be examined as potential pathways for combinatorial accretion — a process where new ideas or concepts are built by merging existing ones. Conceptual development in segregation research (SR) can thus be explored through the co-occurrence of SFs over time. Notably, our findings show that 54% of co-occurrences between two SFs precede their synthesis into a trigram (see ST04 in SI). This is a striking result, as it highlights that the co-occurrence of bigrams likely leads to trigram creation, suggesting that certain pairs of SFs often coexist in discourse or empirical observations before fully merging into more complex trigram constructs. In this section, we will investigate these evolving connections within the literature sample.

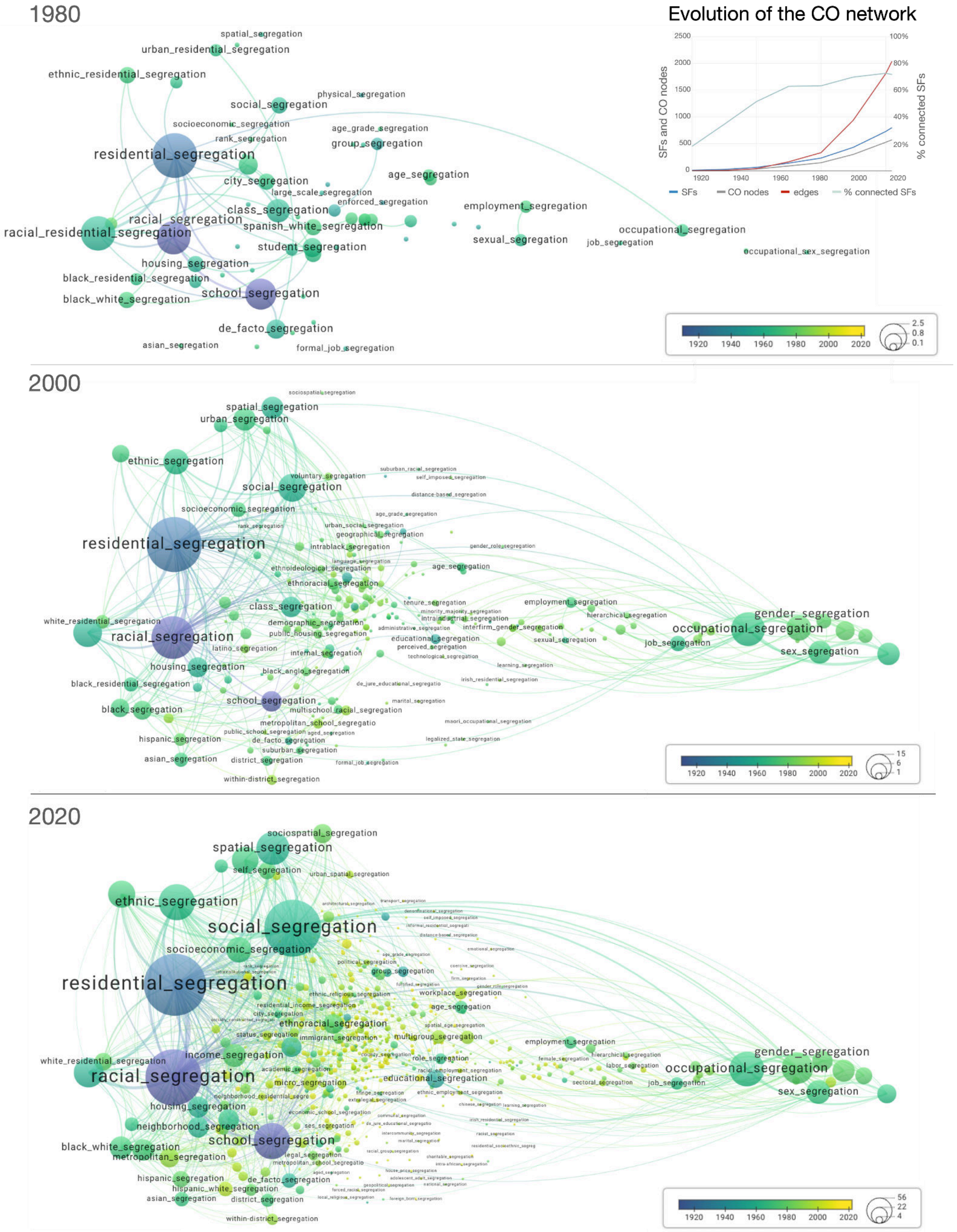

Fig. 5 – New segregation forms and growing connectivity in co-occurrence networks over time. Colors represent time, and node sizes represent degree centrality. See the complete temporalized network, interactive networks per decade[6] and Table ST05 for the 1913-2022 period.

[6] Interactive CO networks of segregation forms for individual decades: 1960, 1970, 1980, 1990, 2000, 2010, 2020 and 2022.

The frequency of COs reflects the intensity of relationships between SFs as authors in the field integrate them. There has been a clear increase in the co-occurrence of SFs over time, indicating a growing trend of authors including more than one segregation form in their work. The average number of SFs per paper rose from 1 in 1960 to 1.4 in 2022 (see Table ST05 in SI). The maximum number of COs in a single paper peaked at 9 in both 2009 and 2020. Figure 5 illustrates how SFs have become increasingly interconnected over time. The increase in SF numbers and connectivity is remarkable from the original triad of racial, school and residential segregation identified in the 1910s and 1920s. In 1980, there were 64 SFs, 33 of which (51.6%) showed connections in the form of co-occurrences. In 2022, there were 804 SFs, and 577 (71.8%) were connected. The number of edges increased more than tenfold, from 53 in 1980 to 6,024 in 2022. Changes between periods are visible, for instance, in the growth of centrality of social segregation and the emergence of the gender segregation and occupational segregation cluster.

**a** The topological landscape of segregation research: main clusters in the co-occurrence network

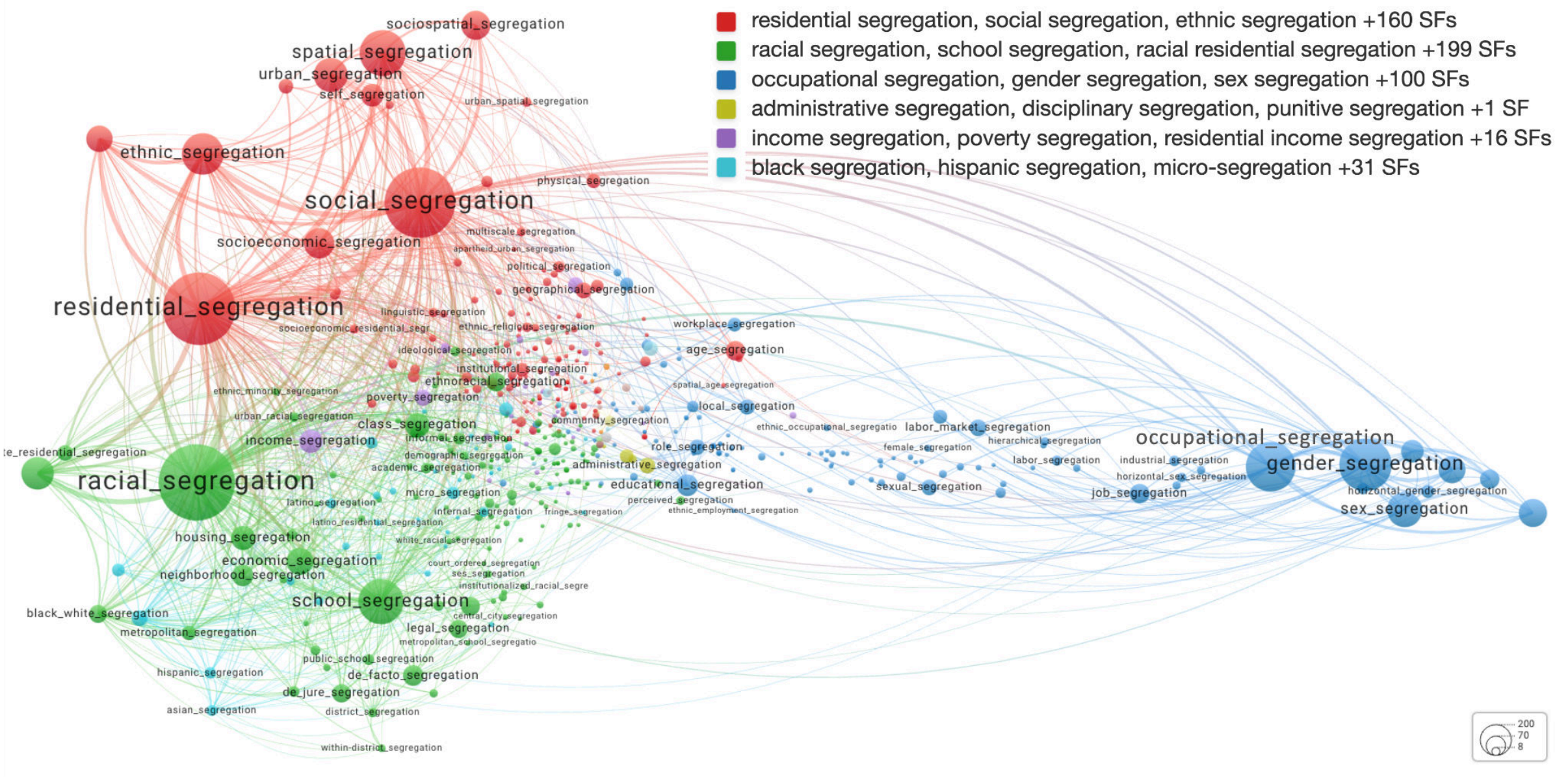

**b** A section of the co-occurrence-based network of segregation forms

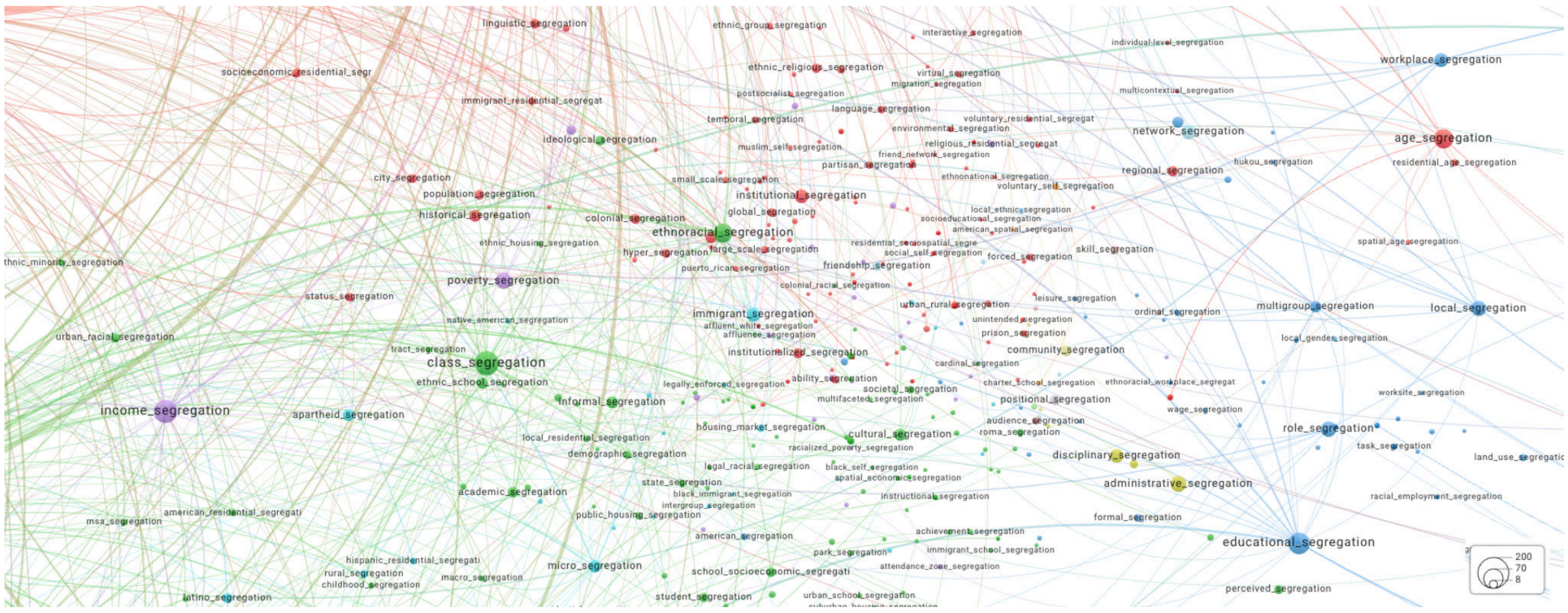

Fig. 6 – The topological landscape of segregation research: (a) The emergent pattern of main SFs, CO-based connections analysed in 22 Louvain clusters with 577 SFs in 2022. Clusters 1-6 contain 537 SFs (93%). SFs listed in the legend are ordered according to degree centrality. Node sizes indicate degree centrality. (b) Zooming in the CO network shows its complexity. Navigate the interactive CO network and see Table ST06.

In addition to hierarchies of betweenness centrality (see Section 6), the complexity of the network can be decoded through a cluster analysis. Louvain clusters are calculated to detect communities based on optimizing the network's modularity, a measure of the density of links inside communities compared to links between communities (Blondel et al., 2008). High modularity indicates a structure with dense connections within communities and sparse connections between communities. Figure 6 shows 22 Louvain clusters as an optimal community aggregation. Three main clusters are visible: the first one (red) relates social segregation and residential segregation, along with ethnic, socioeconomic, urban and spatial segregation; a second one (green) associates racial segregation and school segregation, along with economic, class, legal, institutionalized and state-sanctioned segregation. The third cluster (blue) shows associations of occupational and gender segregation, including vertical, job, sex, women and occupational gender segregation. Educational segregation belongs to this cluster, yet it is a bridge to others, including a smaller one (purple) encompassing income-related forms such as poverty segregation. Figure 7 isolates the three largest clusters.

Co-occurrence network of segregation forms: main clusters

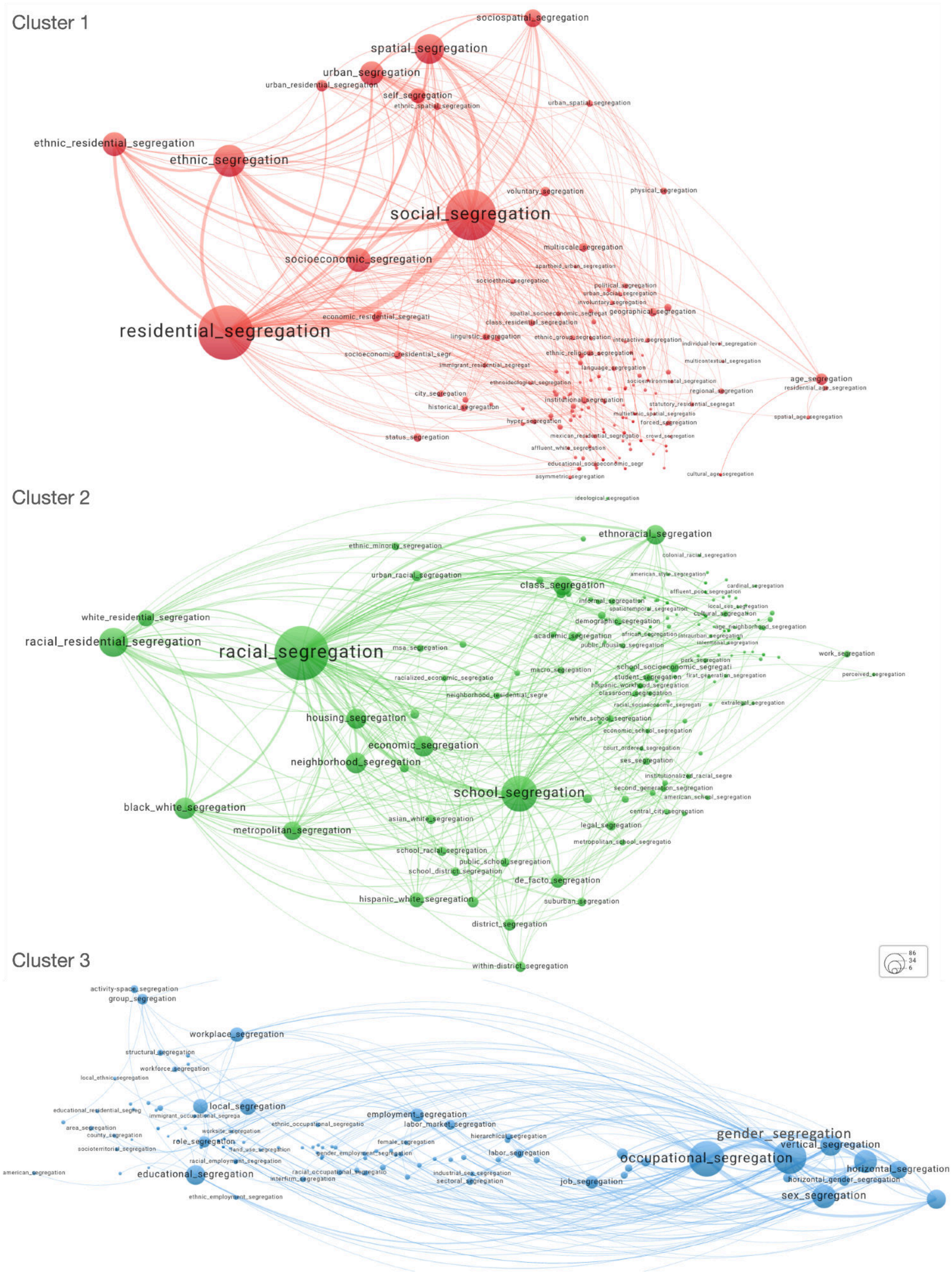

Fig. 7 – The three largest Louvain clusters of segregation forms.
See the interactive CO networks: cluster_1, cluster_2, cluster_3.

Interestingly, these clusters do not amount to exclusive sets of similar SFs – e.g. only with racial and ethnicity-related SFs. Instead, this network is highly centralized around certain SFs (namely, social, residential, racial, school, gender and occupational segregation), bridging clusters (see Section 6). Co-occurrence connections seem frequently based on complementarities between SFs, potentially helping to weave a *transdisciplinary* landscape of segregation research. However, how exactly could SFs transcend disciplinary boundaries?

## 4. Multidisciplinarity, transdisciplinarity and intersectionality in SR

A possibility opened up by mapping the landscape of SR is to assess the role of disciplines as potential borders. A way to do so is to measure the levels of multidisciplinarity and transdisciplinarity in segregation research. These two concepts overlap, so we need to clarify what they imply. Usually, distinctions between them are drawn from the perspective of research work. Multidisciplinarity involves different disciplines independently focused on a common phenomenon. Transdisciplinarity integrates disciplinary-specific conceptual frameworks (Bernstein, 2015; Meeth, 1978; Rosenfield, 1992). From an empirical analysis perspective, we consider multidisciplinarity as the level of diversity of disciplines engaged with segregation phenomena and transdisciplinarity as the extent to which a segregation form arises from within one disciplinary field and subsequently sparks interest across others. Moreover, the social sciences have increasingly recognized that individuals may navigate overlapping social positions such as race, ethnicity, gender, and sexuality, all of which shape their identities (G. R. Bauer et al., 2021; Cho et al., 2013). These approaches, known as "intersectional," are closely linked to issues of segregation. We shall assess intersectionality in segregation research by looking into co-occurrences of segregation forms relating to such positions.

### *4.1. Multidisciplinarity*

Segregation is studied in many disciplines (Fig. 1a; 8b), but *how* multidisciplinary is segregation? What is the level of diversity of disciplines in segregation research?

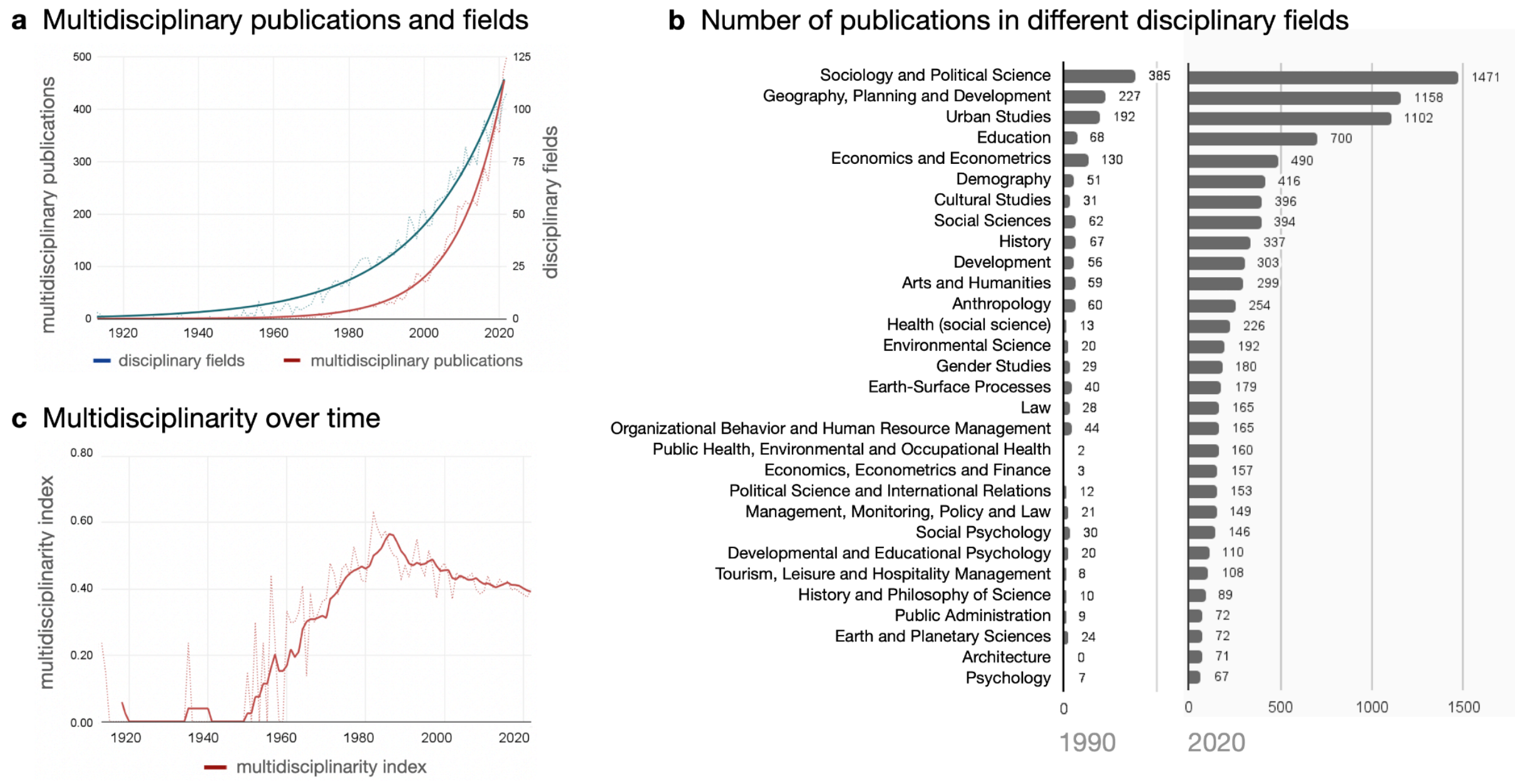

Fig. 8 – Multidisciplinarity in SR: (a) The number of publications in multidisciplinary journals (i.e. journals with more than two ASJC disciplinary fields) (blue line), and the number of ASJC disciplinary fields per year (red line) have grown exponentially. (b) According to the number of publications on segregation in 1990 and 2020 in Scopus, the largest disciplinary fields show increasing concentration. (c) The multidisciplinarity index measures the diversity of disciplinary fields in segregation research over time through Shannon entropy (red line).

We have seen that the Scopus database covers almost 18,000 titles from over 5,000 publishers. Each journal in the database is assigned to one or more subject categories using Scopus' All Science Journal Classification (ASJC) codes. There are 334 ASJC categories or fields. We narrowed our search to 169 disciplinary fields focused on segregation in social phenomena and belonging to the social sciences and humanities (see Table ST07 and Table ST01 in SI). Each SF is published in a document associated with one or more ASJC disciplinary categories. We analyse multidisciplinarity in three levels: (a) multidisciplinarity as the diversity of disciplines in segregation research and (b) the diversity of SFs explored in individual disciplines.

*Multidisciplinarity in segregation research over time*

To assess the diversity of disciplinary fields in SR, we measured (i) the number of fields that published documents on segregation per year, (ii) the number of multidisciplinary publications where SFs appear, and (iii) the diversity of disciplinary fields in SR. The number of fields in SR has been growing exponentially, reaching 101 in 2022 (Fig. 8a, blue line). The number of multidisciplinary publications (i.e. with more than 2 ASJC fields) has increased even more steeply (Fig. 8a, red line). We assessed multidisciplinarity using the same method deployed to assess SF diversity. Using Shannon Entropy as a disciplinary diversity index, we considered the distribution of SFs published yearly over 169 disciplinary fields. Maximum entropy would be reached if the SF distribution was even across the fields. Entropy increased since the 1950s, reaching the peak around 1990, then falling to find stability in the 2010s (red line in Fig. 8c). The falling diversity indicates that the publication of SFs became more concentrated in specific fields, as differences in distributions in 1990 and 2020 show (Fig. 8b).

*Segregation diversity in individual disciplinary fields*

We analysed how individual disciplines have been open to different segregation forms. We measured the diversity of SFs published annually in each ASJC disciplinary field in our sample. SF diversity rose from the 1970s onwards, with Sociology and Political Science leading since then. Geography, Planning and Development and Urban Studies were next from the 1980s onwards, followed by Economics and Education in the 1990s (Fig. 9a). The fact that the average diversity of SFs within disciplines (black line in Fig. 9a) has reached a peak in the late 1980s to subtly fall and become stable since then suggests that disciplines have become more centered around certain SFs.

## *4.2. Transdisciplinarity: segregation forms across disciplinary boundaries*

Many segregation forms transcend original fields. The presence of SFs across disciplines can be interpreted as a trace of their transdisciplinarity. We assess how individual segregation forms progressively transcended disciplinary boundaries by (a) quantifying the changing disciplinary diversity of SFs over time and (b) analysing their co-occurrence network relative to disciplinary fields.

*Evolution in the transdisciplinarity of individual segregation forms*

We measured the Shannon entropy of the distribution of disciplines in which SFs have appeared over time relative to the total number of disciplines. A high Shannon entropy indicates that SFs are not only spreading into more disciplines but are also more evenly distributed across them, reflecting a high degree of transdisciplinarity. The diversity curves in Figure 9b show that SFs have progressively transcended disciplines over time, suggesting an increasing degree of cross-disciplinary integration around them. Some SFs are particularly transdisciplinary. Racial segregation (brown line) shows early signs of spilling over different disciplines since the 1950s, which intensified from the 1970s onwards. Residential segregation (orange line) rapidly grew in transdisciplinarity in the 1970s. School segregation (green line) and social segregation (yellow line) have spread into more disciplines since the 1960s, a trend that intensified in the 1990s. In turn, gender segregation (blue line) emerged and quickly grew in transdisciplinarity from the 1980s onwards. The average degree of transdisciplinarity peak in the late 1980s suggests a slight concentration of SFs around specific disciplines since then.

Spatial, occupational, ethnic, school, gender residential, social and racial segregation forms transcended disciplinary boundaries the most, above the average in 2022 (Fig 9c).

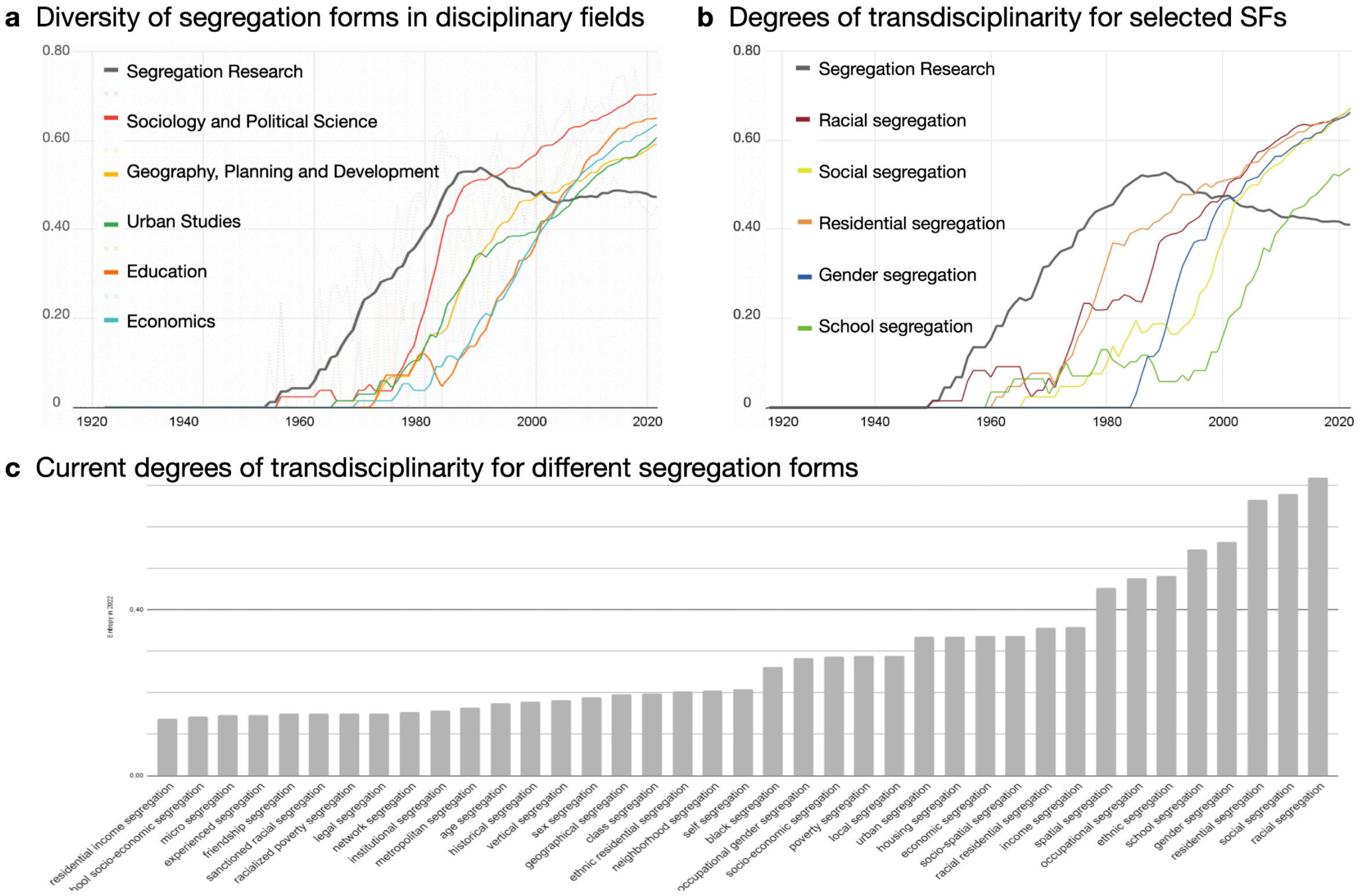

Fig. 9 – (a) Diversity of segregation forms in individual disciplinary fields over time: top 5 disciplines with the most diversity in SFs (colored lines) and the average in SR (dark line). (b) Degrees of transdisciplinarity or disciplinary diversity for selected SFs over time and the average in SR (black line) (moving averages 10 years). (c) Transdisciplinarity in 2022 was measured through Shannon entropy applied to the distribution of SFs across 169 fields. The dark horizontal line is the average degree.

*From original disciplines to disciplines where SFs appear the most*

In addition to the evolution of transdisciplinarity degrees, we assessed transdisciplinarity in the trajectory of SFs over time by looking into changes from their disciplinary origins to the disciplines where they currently most appear. Changes in disciplines from original appearance to a dominant presence signal a transdisciplinary condition. First, we considered the discipline of origin of each SF, i.e. the discipline where each SF was first published, according to the All Science Journal Classification (ASJC). The discipline of origin refers to the field associated with the first publication that contains ASJC information. For simplicity, only the first discipline indexed in the ASJC code field was considered – e.g. social segregation first appeared in the Scopus sample in a journal classified as Sociology and Political Science (Fig. 10a). Second, we considered the degree of presence of SFs in different disciplines in accumulated publications up until 2022. Looking through the lens of SFs, we assessed *the discipline where a segregation form appears most frequently* – e.g. social segregation and racial segregation most appear in journals classified under Geography, Planning and Development (Fig. 10b). To identify the discipline in which SFs appear most frequently, the total number of publications for each SF was counted across all disciplinary fields, based on ASJC codes associated with the journals of publication. The discipline with the largest number of publications was defined as the one where each SF appeared most frequently. In cases where a form of segregation had such a dominant presence in multiple disciplines — i.e. where the number of publications was equal across multiple disciplines — the discipline of origin was used.

Looking through the lens of disciplines, we also assessed *which SFs appear the most in each discipline* (Fig. 10c). We counted which SFs most appear in each disciplinary field – e.g., the SF which most appears in Sociology and Political Science is residential segregation. Three segregation forms (racial, social and residential SFs) correspond to nearly 50% of the forms published in many top disciplinary fields, indicating thematic concentration co-existing with diversification trends.

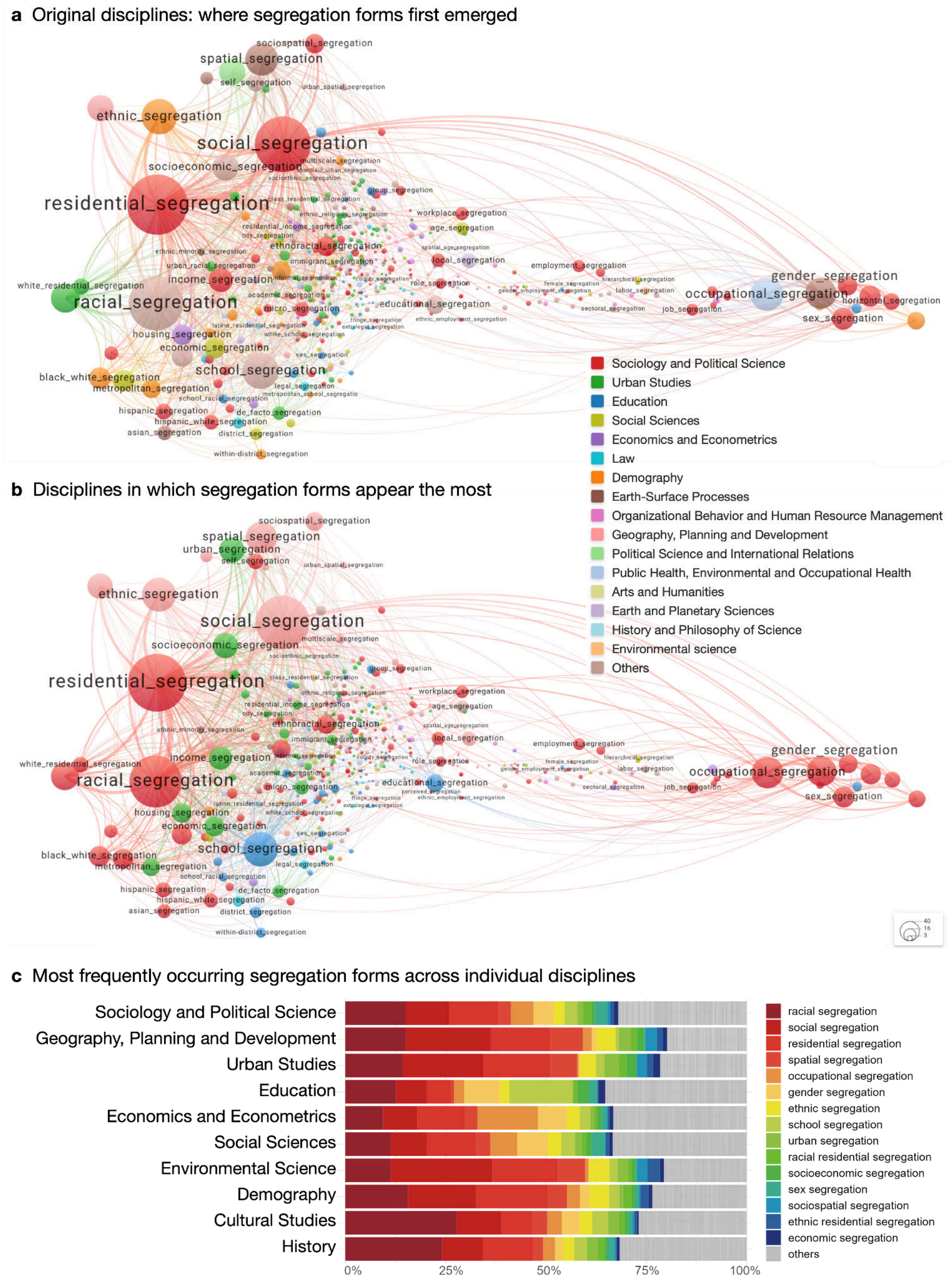

Fig. 10 – (a) Disciplines in which SFs first emerged (Scopus database). (b) Disciplines in which each SF appears the most. Node sizes indicate degree centrality. See the interactive CO networks: [SFs original disciplines](), [disciplines with dominant SFs](). (c) SFs that appear the most in the top 10 individual disciplines in the number of publications. See Table [ST08]() and Graph [G03]().

We examined whether the discipline where SFs had a dominant presence matched their discipline of origin by verifying (1) whether the discipline where an SF has a dominant presence remained the same as its discipline of origin and (2) whether the discipline of dominant presence diverged from the discipline of origin. Two other categories were considered: (3) when a form of segregation was dominant in more than one discipline and (4) when no disciplinary data were available for an SF. Results show that 248 SFs are dominant in their disciplines of origin (30.9%); 100 are dominant in a discipline other than their original ones (12.4%); 374 SFs are dominant in more than one discipline (46,5%); and 82 SFs have no disciplinary information within the dataset (10.2%). Nearly 60% of SFs came to have the strongest presence in disciplines other than their original ones.

A step further, the co-occurrence networks of both origins and dominant presence in disciplines show that SFs strongly interact across disciplines (Fig. 10). In fact, the comparison between these networks with the Lovain clusters network (Fig. 6) suggests that disciplinary borders have not shaped the SF network structure. Considering that segregation is studied in numerous disciplinary fields, this panorama suggests a fluid nature of segregation research, where segregation forms are remarkably connected across disciplines. This is the case despite trends to slight concentration of specific disciplines around certain SFs (Fig. 9a) and vice-versa (Fig. 9b) from the 1990s onwards.

### *4.3. Intersectionality in segregation research*

How have growing concerns with the intersectionality between different social identities in sociological and cultural analysis related to segregation research? We approached this question by identifying documents that combined segregation forms addressing social identities or positions used in quantitative intersectionality analyses: (1) sex/gender/sexual orientation, (2) race/ethnicity, (3) SES (socioeconomic status)/income, (4) age/generation, (5) immigration status and nationality, (6) health and disability, (7) religion/religiosity (cf. Bauer et al., 2021). Segregation forms combining at least two of these positions were considered intersectional SFs (trigrams such as "adult gender segregation" or "racialized economic segregation"). However, these intersectional SFs were relatively infrequent in our dataset. Instead, we primarily focused on the more frequent co-occurrences of SFs that addressed one or more social positions within a single work, which we referred to as "intersectional works" in segregation research (e.g. "racial segregation" and "gender segregation"). The first intersectional co-occurrences in the sample appeared in 1977, combining economic and racial positions (namely, racial residential, socioeconomic residential and class segregation – Farley, 1977). We identified the largest numbers of intersectional co-occurrences of SFs accumulated in 2022 in the Scopus corpus (Fig. 11a). Then we applied Shannon Entropy as an intersectionality index to the distribution of intersectional COs relative to the total number of COs in the Scopus corpus over time (Fig. 11b). There is substantial growth in combinatorial work along intersectional lines in SR. Race/ethnicity is the most frequent position in those combinations.

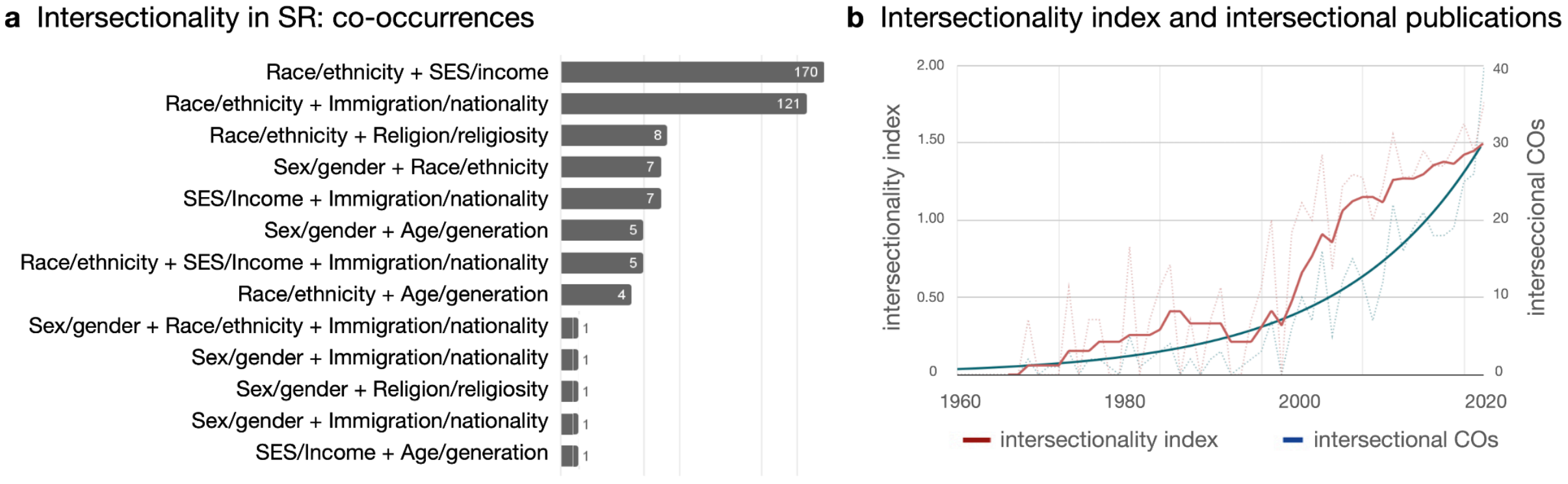

Fig. 11 – (a) Intersectional co-occurrences with the most publications accumulated in 2022 (logarithmic scale).
(b) Intersectionality index based on intersectional co-occurrences measured through Shannon Entropy (red line) and the number of intersectional COs appearing in publications per year (blue line, 10 years moving average).

Our examination of SR through the lens of multidisciplinarity, transdisciplinarity, and intersectionality reveals specific trends in the field. Since the 1990s, there has been a notable increase in the number of disciplines and publications addressing segregation, although more recent years show a concentration of research within fields such as Sociology and Political Science. The transdisciplinary nature of SFs, such as racial and residential segregation, has expanded, with many forms now crossing disciplinary boundaries. However, intersectional research – particularly studies combining multiple social identities like race, gender, and socioeconomic status – remains relatively limited and centered around race/ethnicity. Therefore, while SR demonstrates increasing multidisciplinarity and transdisciplinarity, intersectionality continues to emerge as an area of potential expansion.

## 5. The role of context

What is the role of *context* in identifying SFs? Which segregation forms are researched SFs in different regions of the world? We shall now look into potential geographical specificities in SR. The following contextual trends across continents and countries could be observed.

### *5.1. Presence of continents and countries in SR*

We analysed the research context in SR according to the continent ("world region") and country of institutions of all authors of documents in the Scopus dataset. The proportion of publications at the continental scale (see Fig. 1b in the Introduction) shows North America's significant but diminishing presence relative to other regions (Fig. 12a).

While academic production in Asia has steadily grown since the 1970s, European production has manifolded. Production in Africa and Oceania has been comparatively small but stable since the 1980s, while South America has risen to similar percentages in the 2000s. At the country scale, the initial academic attention to segregation in the US consolidated into the largest proportion of publications in the Scopus dataset (Fig. 12b). However, there has been a trend of reduction in this proportion since the 1980s. Production has substantially diversified (see "other countries" in Fig. 12b) and increased in countries like Spain, China and the Netherlands. In sheer numbers, the US and the UK have the largest publication volumes in SR, which is plausible given our focus on English-language publications (Fig. 12d).

### *5.2. Most researched segregation forms across world regions*

We analysed the number of publications with the most frequent forms of segregation per country. Many SFs have been identified across different countries. The top SFs are likely to be relevant subjects in all continents (SFs in red in Fig. 12c) – most notably, social segregation (the most researched SF in four out of six continents), racial segregation (top 1 in two continents and top 5 in other three) and residential segregation (top 2 in five continents and top 5 in the remaining continent). There is considerable variation across continents, most visible near the bottom half of the table (top 9-20 SFs in Fig. 12c), suggesting emphases that are specific to certain regional contexts or SFs less likely to be recognised as top research issues in every region. In **North America**, racial segregation has the strongest presence, followed by residential, social and school segregation. Racial issues can also be seen in racial residential, de facto and black-white segregation (top 7, 8 and 20). In **Europe**, social, gender, ethnic and spatial segregation find higher positions. **Asia** shows similar trends, with occupational segregation taking the 4th spot, followed by spatial and racial segregation. Apart from the top universal SFs and having racial segregation as its most frequently published SF, **Africa** shows specific SFs centered on the legally enforced status of racial segregation, with apartheid segregation as the 6th most published one, followed by racial residential (top 9), informal (11), institutionalized racial (14), urban racial (15), colonial (17), historical (18) and de jure (20) segregation forms. In **South America**, the segregational effects of social and spatial inequalities are felt in the presence of SFs such as spatial, sociospatial and urban segregation; socioeconomic, socioeconomic residential segregation; and so on. Context is felt in territorial (15), poverty (16) and physical (19) segregation forms. **Oceania** shows a concern with gender and employment-related SFs, including intersectional ones, in top positions, namely gender (top 4), occupational (5), sex (9), occupational sex (10), occupational gender (13), and vertical segregation (12) SFs.

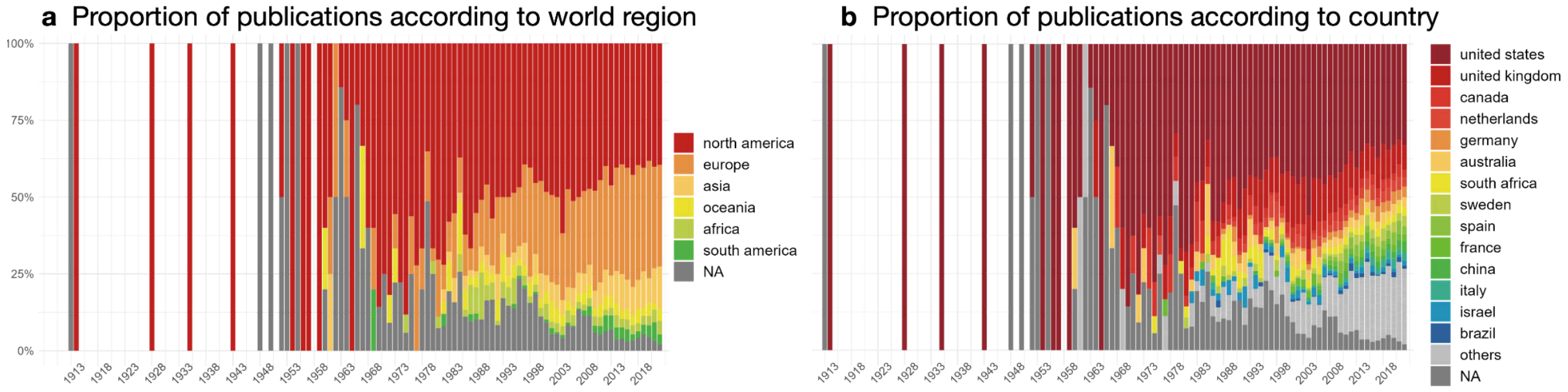

**c** Regional trends: most frequent segregation forms in different world regions

| | North America | Europe | Asia | Africa | South America | Oceania |
|---|---|---|---|---|---|---|
| 1 | racial segregation | social segregation | social segregation | racial segregation | social segregation | social segregation |
| 2 | residential segregation | residential segregation | residential segregation | residential segregation | residential segregation | racial segregation |
| 3 | social segregation | gender segregation | gender segregation | social segregation | spatial segregation | residential segregation |
| 4 | occupational segregation | racial segregation | occupational segregation | urban segregation | sociospatial segregation | gender segregation |
| 5 | school segregation | ethnic segregation | spatial segregation | spatial segregation | urban segregation | occupational segregation |
| 6 | gender segregation | spatial segregation | racial segregation | apartheid segregation | socioeconomic segregation | ethnic segregation |
| 7 | racial residential segregation | occupational segregation | ethnic segregation | occupational segregation | school segregation | spatial segregation |
| 8 | spatial segregation | school segregation | sociospatial segregation | gender segregation | racial segregation | ethnic residential segregation |
| 9 | sex segregation | urban segregation | urban segregation | racial residential segregation | gender segregation | sex segregation |
| 10 | ethnic segregation | sociospatial segregation | sex segregation | socioeconomic segregation | educational segregation | occupational sex segregation |
| 11 | economic segregation | socioeconomic segregation | socioeconomic segregation | informal segregation | ethnic segregation | school segregation |
| 12 | occupational sex segregation | ethnic residential segregation | school segregation | class segregation | occupational segregation | vertical segregation |
| 13 | income segregation | occupational gender segregation | occupational gender segregation | ethnic segregation | socioeconomic residential segregation | occupational gender segregation |
| 14 | housing segregation | sex segregation | self segregation | institutionalized racial segregation | self segregation | socioeconomic segregation |
| 15 | class segregation | vertical segregation | educational segregation | urban racial segregation | territorial segregation | sociospatial segregation |
| 16 | socioeconomic segregation | horizontal segregation | occupational sex segregation | sociospatial segregation | poverty segregation | self segregation |
| 17 | neighborhood segregation | self segregation | economic segregation | colonial segregation | labor segregation | educational segregation |
| 18 | urban segregation | occupational sex segregation | class segregation | ethnic residential segregation | school socioeconomic segregation | group segregation |
| 19 | de facto segregation | educational segregation | geographical segregation | territorial segregation | physical segregation | community segregation |
| 20 | occupational gender segregation | class segregation | housing segregation | de jure segregation | economic residential segregation | physical segregation |
| | range: from 1368 to 54 papers | range: from 850 to 32 papers | range: from 208 to 8 papers | range: from 168 to 3 papers | range: from 75 to 3 papers | range: from 73 to 3 papers |

**d** Number of publications according to country

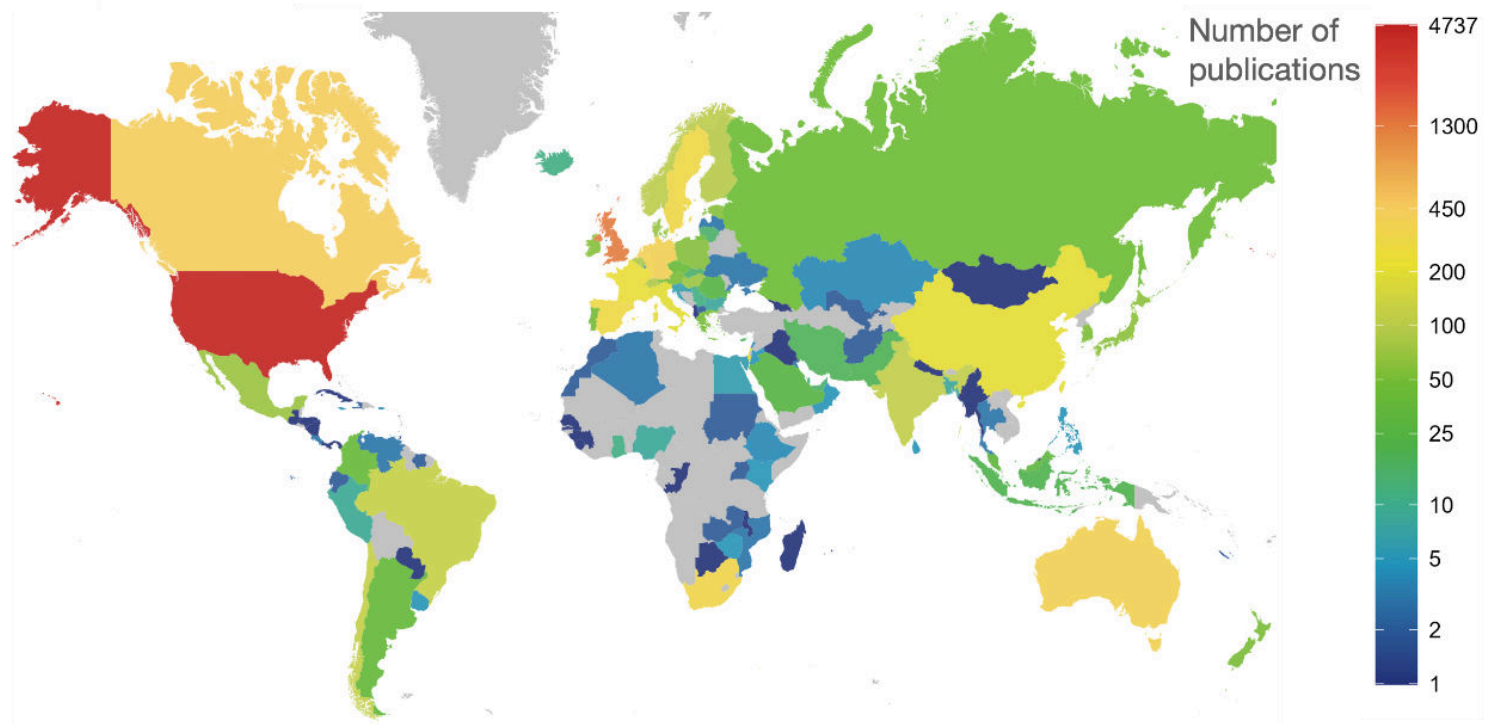

| Legend | Segregation form | World regions |
|---|---|---|
| | appears in all | six out of six |
| | appears in all but one | five out of six |
| | appears in four | four out of six |
| | appears in three | three out of six |
| | appears in two | two of six |
| | appears in only one | one of six |

Fig. 12 – (a) Proportion of publications in SR according to world regions or continents (Table ST09 and Graph G04 in SI). (b) Proportional of publication in SR across countries (Table ST10 and graph G05). (c) Regional trends and the twenty most frequent SFs in different continents, with the number of publications (range) from the top 1 SF to the top 20 in each continent: SFs in red appear in all continents; in turn, SFs in blue appear in only one continent. (d) Distribution of the total number of publications in SR in the period within the Scopus database across 127 countries.

## 6. Path dependence: the historical shaping of the field

We have seen that place matters in segregation research. Along with certain generally dominant forms, there are regional differences in the panorama of segregation forms. Would time also matter? What is, if any, the weight of the preceding history of developments in this heterogeneous field? Could the paths taken in the field affect the identification of new SFs? We shall examine these questions through the concept of *path dependence* – the idea that processes and outcomes are shaped and constrained by the sequence of previous events, often locking in specific paths due to past events or developments (Arthur, 1994; David, 1985). Indeed, a scientific field can take several possible paths and find distinct outcomes in the long run. Empirical problems may be explained in more than one way. Conceptualisation can go in different directions. These possibilities complicate mapping the evolution of SR in its genealogical paths from one segregation form to another. However, epistemologically, path dependence implies that the space of possible ideas is not explored ergodically (cf. David, 2007), meaning that researchers cannot explore every possible concept or combination in their search for understanding segregation forms. Instead, concept formation and empirical discoveries will likely be built upon previous empirical experiences and cognitive tools. Kuhn (1970) famously observed that a particular framework may shape how future research questions are asked, evidence is interpreted, and phenomena are recognized. The initial choices about which theories or methods to adopt might influence future discoveries and knowledge production. Existing SFs in the literature may guide interest in certain ranges of possible forms of segregation and influence future research. We hypothesize a few ways in which path dependence can play a role in segregation research:

(i) *Phenomenal visibility*. Certain SFs may be socially and materially more apparent than others – say, the radical racial segregation in the US or South Africa or physically segregated residential areas in major Brazilian cities, probably easier to spot than, say, the segregated movement of socially different people in the streets. That said, it does not follow that there is a deterministic relationship between phenomenal visibility and early conceptual recognition. For instance, gender segregation took a considerable time to appear in the segregation literature, and did so in the context of occupational segregation research (e.g. Harkness & Super, 1985; see Lorber, 2010).

(ii) *Epistemological visibility.* In line with Kuhn's pioneering observation, well-known problems and approaches might influence the selection of a particular research subject. If this is the case, the current proportions of existing SFs may increase the probability of certain SFs being selected when researchers produce new works at a given time. A new publication in SR may be more likely to be added to SFs with a high proportion of publications. This flux would coexist with the natural trend to increase diversity in SFs. The falling rate of diversity growth in SR since the 1990s (Fig. 1c; 3c) suggests that time is a crucial condition for segregation forms to capture the attention of researchers in their fields, leading to reinforcing mechanisms in research.

(iii) *Theoretical attention* to certain phenomenal areas or conceptualizations might come at the expense of recognizing others. For instance, the focus on the slow temporality of the production of segregated residential areas established since the Chicago School was almost naturally disassociated from attention to real-time, volatile segregation forms felt in people's daily experiences (Netto et al., 2015; cf. Wissink et al., 2016). Furthermore, the political debates of the times also focus on specific topics, such as racial segregation and the progressive expansion of debates towards ethnicity (see De Lepervanche, 1980; Rodríguez, 2000).

(iv) *Conceptual development*. Combinatorial work seems to constitute the bulk of concept creation (Simonton, 2010; section 3 above). As combinatorial accretion involves existing SFs, it has the effect of asserting their existence, even if adding steps beyond them. An initial set of SFs and concepts may retain foundational value for thinking of new segregation forms.

(v) *Contingency and context*. Context-dependent choices can set in motion sequences of works focused on certain SFs. For instance, the place where the SF is first identified may influence its centrality in the field over time. Early concerns about residential segregation related to race in the US (e.g. Burgess, 1928) might have been influential over the course of emphases worldwide. However, the context-dependency of SFs is frequently forgotten (Brun & Chauviré, 1983). Contextual forces, such as power dynamics, public debates, institutional frameworks, funding priorities, and ideological biases—among others—can shape the segregation research agenda, reinforcing certain paths while limiting others. For example, these forces may create blind spots, where certain forms of segregation are systematically underexplored, or they may prioritize forms that align with political or funding interests. Consequently, the field's development is not solely driven by empirical discovery but may be shaped by these influences, which steer attention toward specific segregation forms and methods, ultimately shaping the trajectory of future research.

These factors, along with other possible influences, may trigger path dependencies, potentially locking the multidisciplinary field into specific trajectories centered around certain forms of segregation. Building on our topological perspective of the landscape of segregation forms identified in the literature, and how this landscape evolves through combinatorial and exploratory accretion processes, these factors could feedback into the dynamics of *preferential attachment* in scientific production (Barabási & Albert, 1999; Peng, 2015). Preferential attachment refers to the process by which nodes in a network are more likely to form links with nodes that are already well-connected. The probability of a new node connecting to an existing node increases with the degree (number of connections) of the existing node. In the context of this study, we empirically identified relationships among SFs through co-occurrences in the literature, generating a network representation. SFs with a higher degree (number of connections) in the network are more likely to attract new links, creating a small number of highly connected nodes. Figure 13c illustrates the distribution of topological centrality across SFs in this network. Centrality was measured using Freeman's (1978) definitions: Degree centrality refers to the number of direct connections a node (SF) has with other nodes, represented by the number of co-occurrences between an SF and other SFs. Betweenness centrality measures how often a node lies on the shortest paths between all other nodes, indicating how frequently an SF serves as a bridge in the network. A more central position indicates that an SF plays a key role in mediating many co-occurrence connections between other SFs.

We proposed an accretion model to explain the dynamics of identifying segregation forms (section 2), which predicts exponential growth. However, the model does not necessarily lead to the highly hierarchical network observed in Figure 13c, where a few SFs show high centrality. The reason may be explained by path dependence in SF identification. Time matters, as SFs identified earlier may serve as anchors, triggering both combinatorial and exploratory accretion and continuing to co-occur with newer SFs over time. As SFs are added through these processes, they foster connectivity among existing SFs and facilitate the identification of new ones. In this context, the properties of networks and their corresponding measures offer valuable explanatory insights. Degree centrality captures local connectivity. In contrast, betweenness centrality captures the larger-scale structure of the CO network. The measure demonstrates that as co-occurrence relationships are considered over time, certain SFs attain privileged positions, linking pairs of SFs throughout the evolving network of segregation forms. Both measures are complementary, as they capture local and global dynamics within the network, highlighting the role of SFs in mediating sequences of co-occurrences produced by the collective work of authors. This analysis suggests that the SF network likely follows a power-law distribution, where a small number of nodes (SFs) have a disproportionately high number of connections, while most nodes have relatively few. This pattern aligns with a preferential attachment dynamic (Barabási & Albert, 1999; Abbasi et al., 2012). Path dependence emerges as the network grows and new SFs are likely to connect and combine with highly connected SFs, reinforcing inequalities in degree and betweenness centrality.

We have seen that 91.1% of trigrams containing words from existing bigrams emerged after the bigrams, and 54% of trigrams appeared after the co-occurrence of their composing bigrams. To empirically test the path dependence hypothesis in SR, we examined the role of time in shaping the connectivity of SFs within the CO network. The path-dependence hypothesis suggests that SFs may acquire centrality over time: SFs identified earlier are more likely to gain centrality as the network grows. We assessed this by analyzing the correlation between the time since SFs first publication and their centrality in the co-occurrence network (Fig. 13, Table 2).

Spearman Correlations between publication time and betweenness centrality are moderate ($R=0.53$, $p\text{-value}<0.01$) but do suggest a degree of path dependence in the centrality of SFs in the network (Fig. 13b). As the field is substantially built on existing SFs, the importance of certain SFs is likely to increase over time. That would be a subtle general trend – which, however, does not explain the high centrality of SFs that emerged much later in the literature, such as gender, occupational and ethnic segregation forms (respectively the 5th, 6th and 8th most central in CO network).

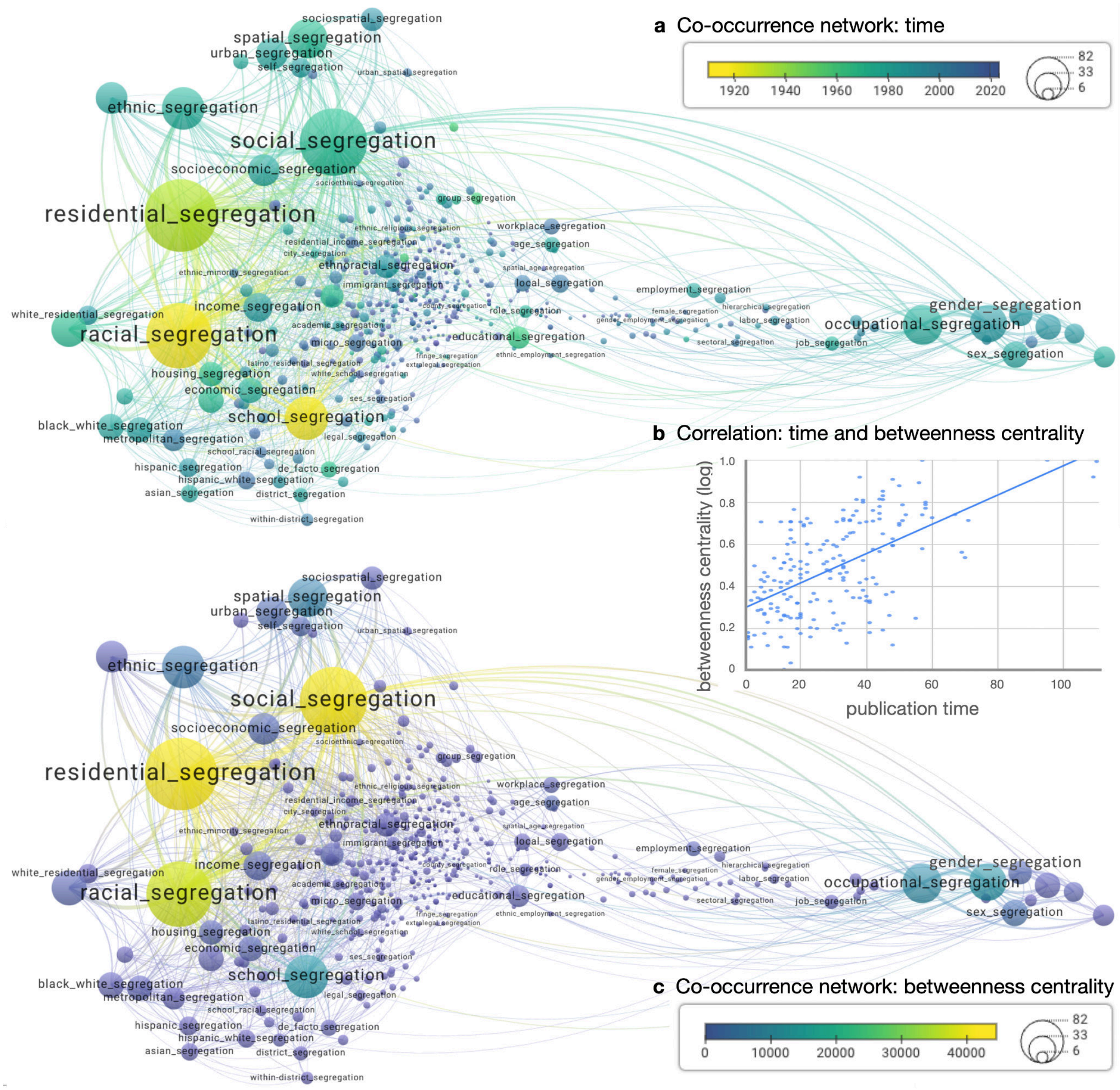

Fig. 13 – (a) SF co-occurrence network based on publication time (first appearance of segregation forms, frequency > 1) from the 1910s (yellow) to 2022 (purple). (b) Spearman correlation between publication time (in years) and betweenness centrality of SFs in the CO network (Spearman correlation R=0.53, p<0.01). (c) SF co-occurrence network based on betweenness centrality (node degree > 1). Node sizes indicate degree centrality. See the interactive networks: time and betweenness centrality.

Is this relative trend of early SFs remaining central in the research landscape contradictory to the dynamics of growth and diversification in SF detected above? There are reasons not to believe this possibility. Even if this particular correlation carries traces of causation, path dependence does not mean the impossibility of change and innovation – even if the field of possibilities seems more constrained to those operating in it. The trend for many SFs to retain centrality over time coexists with the exponential addition of new SFs and seems at work in the non-exponential growth in diversity, including its diminishing rate since the mid-1980s (Fig. 1c), as new publications are also distributed along existing SFs. So, there is room for path dependence *and* conceptual innovation in segregation research. For instance, racial segregation, school segregation and residential segregation are the earliest SFs in the Scopus database and are among the four most central ones. Social segregation appeared much later (1966) to become the second most central SF in the network, virtually tied to residential segregation. Gender segregation appeared in 1985 to quickly become the 5th most central SF. Digging deeper into how SF CO networks unfold over time can reveal clearer traces of path dependence. We can assess path dependence in SR by examining the time and location of the first publications on different segregation forms and their topological centrality within the CO network.

| | Segregation form | First year of publication | Degree centrality | Betweenness centrality |
|---|---|---|---|---|
| 1 | residential_segregation | 1928 | 777 | 44,492.09 |
| 2 | social_segregation | 1966 | 667 | 44,298.05 |
| 3 | racial_segregation | 1913 | 631 | 40,524.88 |
| 4 | school_segregation | 1914 | 277 | 13,447.18 |
| 5 | gender_segregation | 1985 | 205 | 13,307.09 |
| 6 | occupational_segregation | 1975 | 233 | 11,414.06 |
| 7 | spatial_segregation | 1967 | 228 | 8,966.18 |
| 8 | ethnic_segregation | 1973 | 267 | 7,181.67 |
| 9 | sex_segregation | 1978 | 102 | 4,902.72 |
| 10 | urban_segregation | 1978 | 147 | 3,555.43 |
| 11 | socioeconomic_segregation | 1979 | 147 | 2,785.40 |
| 12 | vertical_segregation | 1986 | 81 | 2,708.74 |
| 13 | age_segregation | 1978 | 35 | 2,692.72 |
| 14 | self_segregation | 1984 | 58 | 2,594.48 |
| 15 | income_segregation | 1984 | 124 | 2,234.23 |
| 16 | racial_residential_segregation | 1965 | 180 | 2,225.46 |
| 17 | housing_segregation | 1965 | 88 | 1,889.39 |
| 18 | economic_segregation | 1972 | 93 | 1,856.60 |
| 19 | employment_segregation | 1979 | 33 | 1,651.58 |
| 20 | class_segregation | 1965 | 78 | 1,456.43 |

Table 2 – First year of publication in the Scopus dataset and measures of centrality in the network of co-occurrent segregation forms. See the complete Table ST11 in the SI.

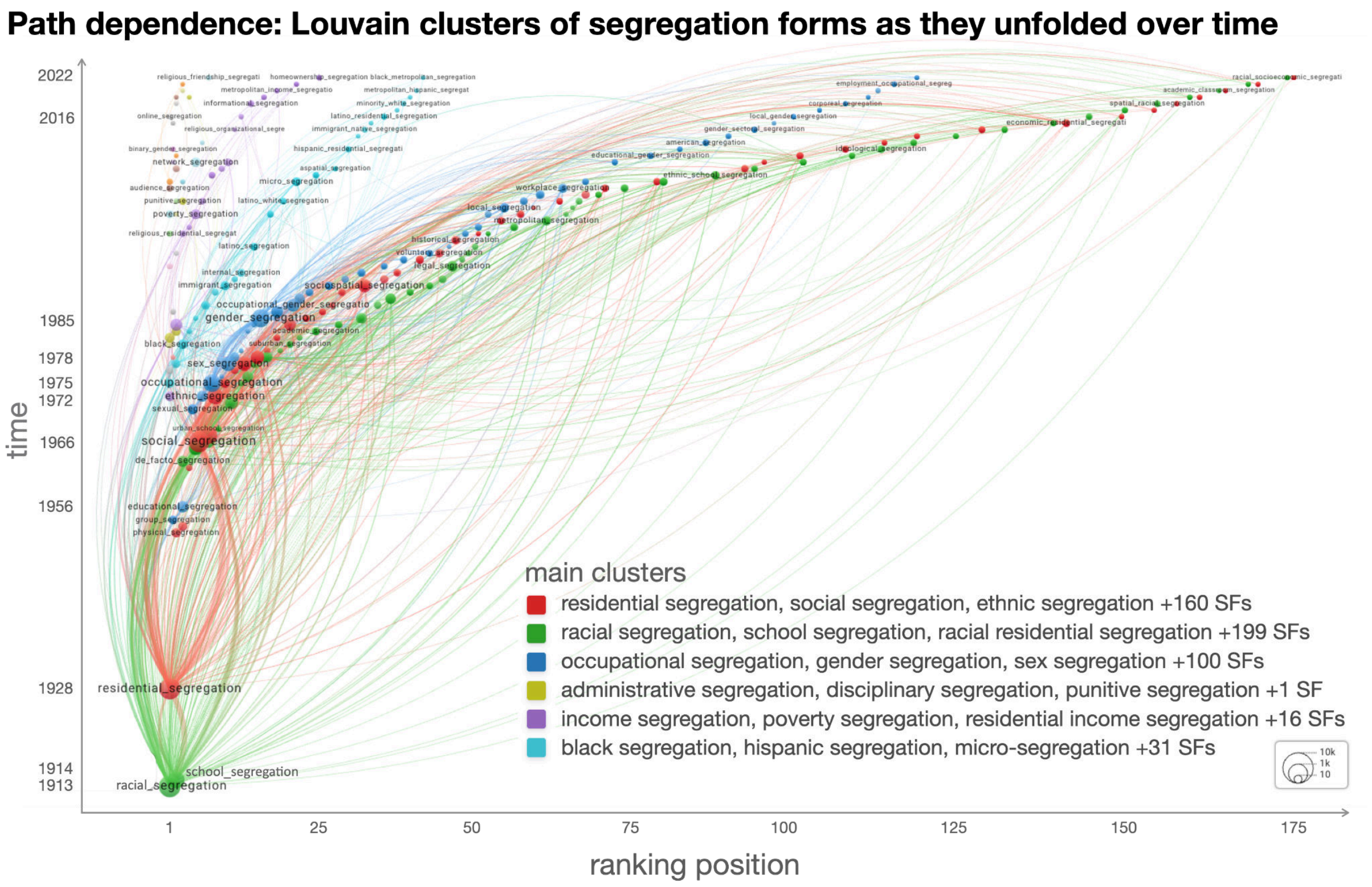

Fig. 14 – Connectivity in time: Segregation forms according to first publication year (Scopus database) arranged in Louvain clusters as they unfold and connect over time in the SF co-occurrence network. The ranking position is defined by the SFs' year of first publication within their specific clusters. Navigate the interactive path-dependence network (select color: clusters).

The earliest segregation forms (SFs) identified in the Scopus dataset—racial, residential, and school segregation—paved the way for other clusters and their interconnections. Tracing how these branches evolved through co-occurrences provides insight into the development of segregation research, including the initial links between SFs in the Scopus sample (Table ST12 in SI). To illustrate this process, we examine two specific examples. Occupational segregation, which first appeared in 1975 (light blue cluster in Fig. 14), co-occurred with racial segregation (1913) for the first time in 1994, with social segregation (1966) in 1985, sexual segregation (1971) in 1981, and ethnic segregation (1973) in 2006, among others. As the field evolved, occupational segregation also formed connections with SFs that emerged later, including sex segregation (1978) in 1988, employment segregation (1979) in 1989, industrial segregation (1989) in 1989, gender segregation (1985) in 1990, vertical segregation (1986) in 1992, and occupational gender segregation (1987) in 1987. Similarly, activity-space segregation (first recorded in 1999) connected with social segregation (1966) and socio-spatial segregation (1990) in 2019. By 2022, it had formed links with spatial segregation (1967), residential segregation (1928), educational segregation (1956), employment segregation (1979), local segregation (2002), racial segregation (1913), and workplace segregation (2005). These co-occurrences suggest that connections between forms of segregation were explored at distinct moments in the field's trajectory, revealing a sequence of conceptual intersections over time.

## 7. Knowledge production networks

We have analysed so far the structure and evolution of the multidisciplinary field of segregation research. However, how is knowledge produced in SR? Now we look into how works relate to each other as they are created, what are the main production sites country-wise, the landscape of collaboration, and the topology of journals publishing works on the multifaceted subject.

### *7.1. Co-citation network of works in SR*

The relationship between different works on segregation can be grasped by how they cite existing ones. A way of doing this is to consider the collection of works cited together by documents in a corpus ("co-citation"). This differs from a direct citation network, which connects citing papers to cited papers and is useful for tracing the chronological development of ideas. The co-citation network is undirected and built through database documents with common references. Two papers are said to be co-cited if at least one document cites both of them. The strength of the relationship between works is indicated by the number of times they are cited together (Osareh, 1996). The approach can capture networks of related documents. Each node represents a document, and an undirected edge between nodes A and B is formed if one or more other documents cite together both documents A and B. This implies that the network analysis goes beyond works in the Scopus corpus to cover 473,706 references – an x-ray of the field. For visualization purposes, we adopted ten as a minimum number of co-citations, i.e. the network shows only works which are co-cited ten or more times by the documents in the sample, with a limit of 1,000 most co-cited documents (i.e. nodes with the highest total link strength or the sum of the number of times the reference appears cited with other works) reduced to 998 after references were checked, and 2,000 visible links (Fig. 15). Node sizes indicate the citation number.

Works were then grouped in six clusters based on network modularity or the density of links inside communities compared to links between communities, calculated through the “smart local moving algorithm” (Waltman & Van Eck, 2013). Clusters show nodes with internal thematic affinities. However, research themes are not exclusive to specific clusters. Cluster 1 (red) is the largest and encompasses works on social, ethnic and residential segregation, including emphases on method and measurement. The article "The dimensions of residential segregation" (Massey & Denton, 1988) is the most cited in this cluster (followed by Duncan & Duncan, 1955), and in the sample as a whole. Cluster 2 (green) concentrates on works related to schools, race, intergroup contact, homophily, social networks and measurement. "The nature of prejudice" (Allport, 1954) is the most cited work. Cluster 3 (dark blue) brings works on urban segregation mostly related to poverty and race. The article "When work disappears: the world of the new urban poor" (Wilson, 1996) is the most cited document. Cluster 4 (ochre) includes a mix of works on policy, race, neighborhoods and ghettos, including the articles "The truly disadvantaged" (Wilson, 1987) and "American apartheid: segregation and the making of the underclass" (Massey, 1990), along with dynamic models of residential segregation since Schelling (1969). The most visually identifiable cluster is 5 (purple), including works on gender, sex, occupational segregation and their intersections with race, education. Cluster 6 (light blue) includes racial and ethnic features. The book "American apartheid" (Massey & Denton, 1993) is the most cited work. The full list of documents with clusters and publication data is available in Table ST13.

**Co-citation network of documents in segregation research**

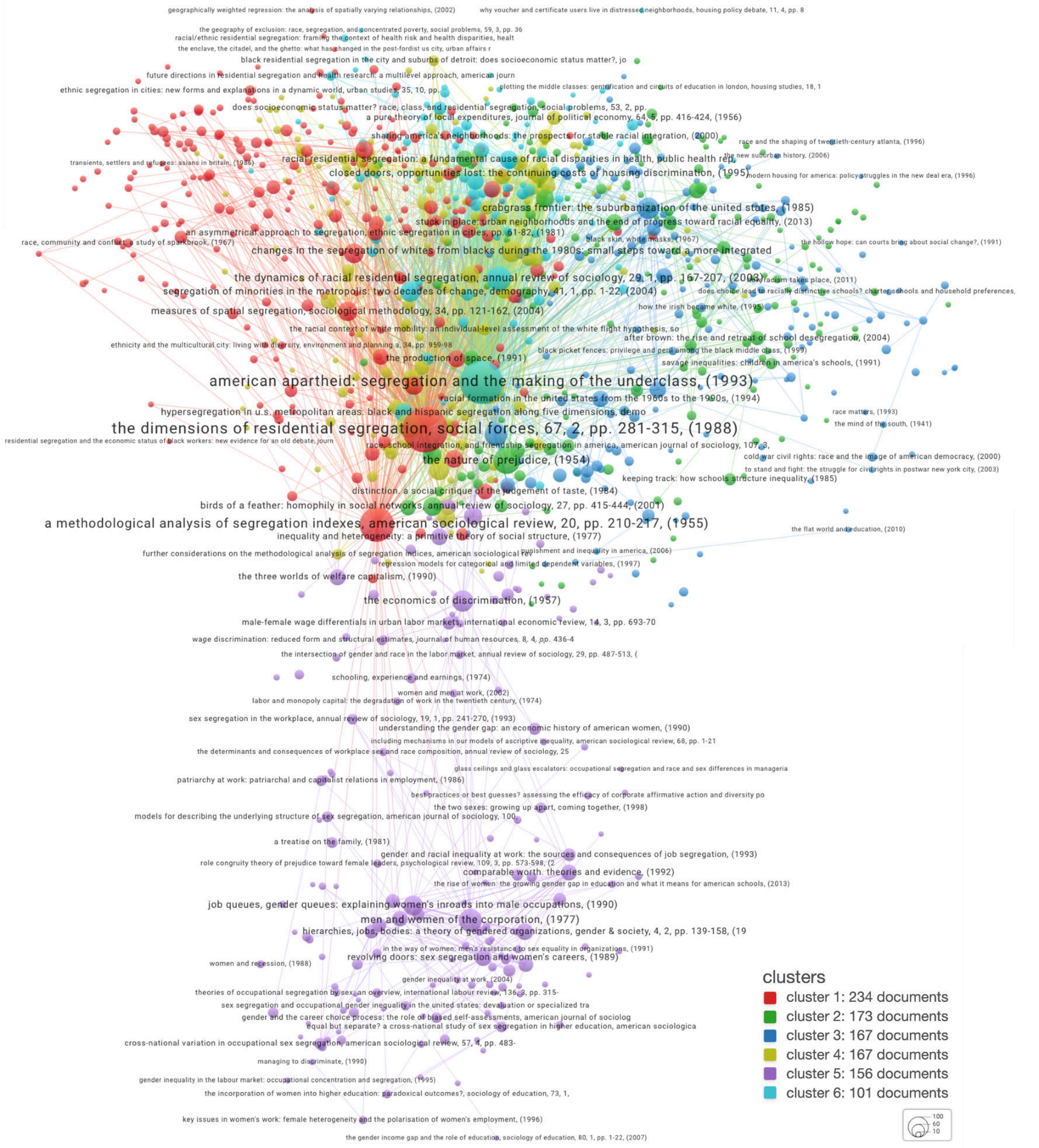

Fig. 15 – Co-citation network of documents in SR. This representation brings the most co-cited 998 documents within the network. Node sizes represent the number of citations. For a comprehensive view, refer to Table ST13 in SI with clusters and the interactive network (layout parameters: attraction 8, repulsion 3, rotate 90 degrees; 2,000 visible links).

### *7.2. Co-authorship network across countries*

We analyzed the field as an international collaboration network using a specialized type of bibliometric analysis where nodes represent countries, and edges represent co-authorship relationships between authors from different countries, identified through author affiliations. The node size (degree centrality) shows the number of unique documents in which authors from a particular country collaborate, providing insight into hierarchical structures. The weight of an edge is determined by the number of co-authored publications between authors from the two countries, reflecting the strength of collaboration. For visualization purposes, the minimum number of documents associated with a country is five. Betweenness centrality (see [network] and Table [ST14] in [SI]) confirms the enormous presence of researchers working in the US context, followed by the UK and a block of countries including Australia, Germany, Netherlands and France. Clusters show which countries appear more frequently together in the collaboration network, and are based on network modularity (Fig. 16a).

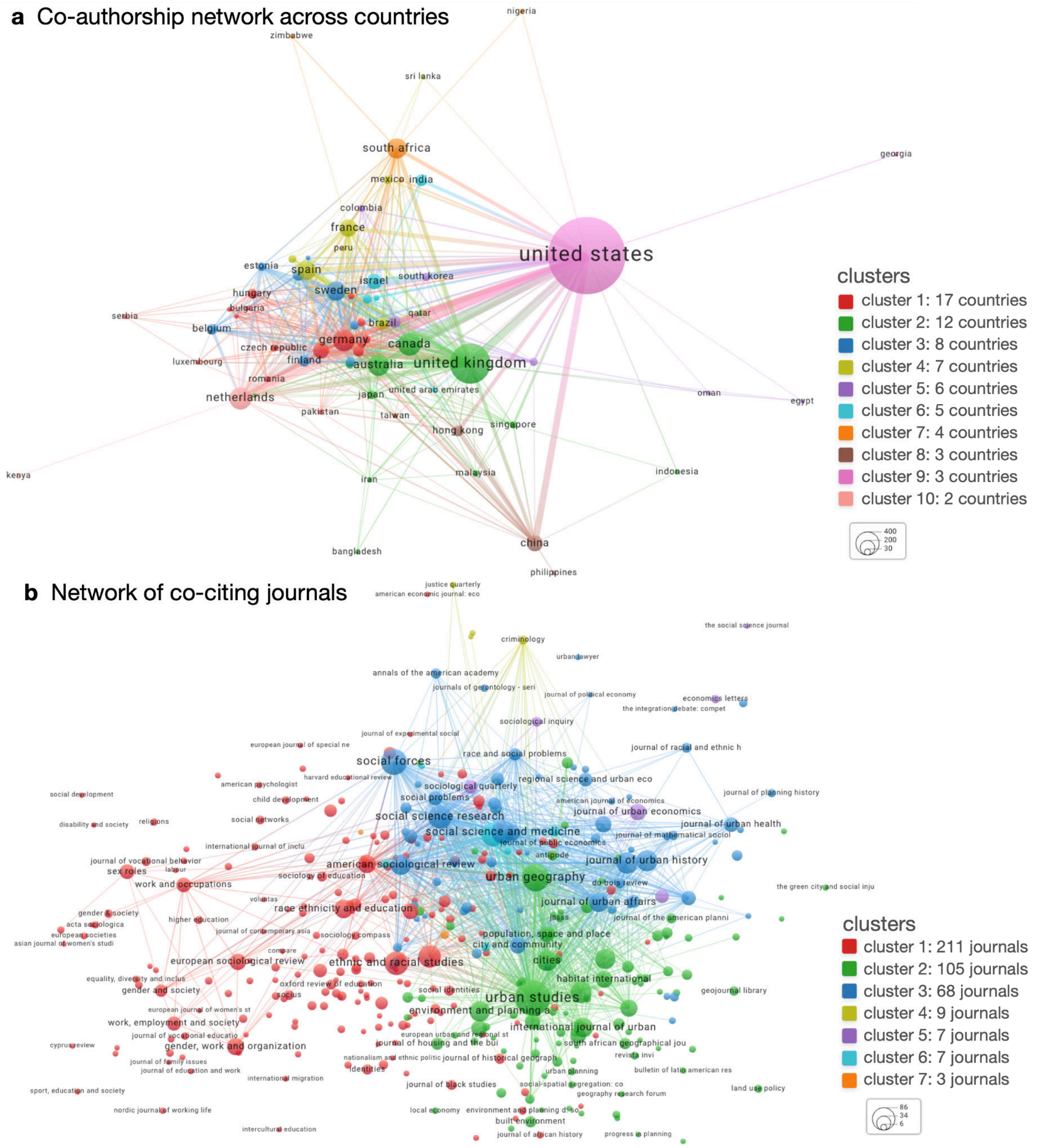

Fig. 16 – (a) Co-authorship network across countries (see the [interactive network] including [betweenness centrality]; layout parameters: attraction 6, repulsion 4, straight links). (b) Bibliographic coupling. Node size indicates the number of documents on segregation published by the journal ([interactive network] including [betweenness centrality]; attraction 5, repulsion 3).

Regarding world region clusters, they are internally diverse, showing that distance is frequently overcome in co-authorship. Cluster 1 (red) includes central European countries such as Germany, Italy, Switzerland, Eastern European countries (e.g. Hungary, Slovakia, Bulgaria), Argentina, Pakistan and Taiwan. Cluster 2 (green) includes UK, Australia and Canada as English-speaking countries, Japan, Singapore and Malaysia. Cluster 3 (dark blue) includes Sweden, Norway, Finland, Belgium and Poland; Cluster 4 (ochre) includes France, Spain and Brazil (see Table ST14 in SI).

### *7.3. Co-citing journals*

When applied to journals, bibliographic coupling focuses on identifying those that share sources, revealing similarities or complementarities in their research topics and focus. Journals A and B are bibliographically coupled if they cite the same journal or set of journals. The strength of the bibliographic coupling is determined by the number of shared references between the two journals. Such connectivity generates a network where each node represents a journal. An edge between two nodes represents a bibliographic coupling relationship. The weight of an edge is determined by the number of references shared between the two journals, reflecting the strength of their bibliographic coupling. Figure 16b shows the top 1,000 journals with the greatest total link strength. The minimum number of citations of a journal considered was five. Node size indicates the number of documents on segregation published by each journal.

The three main clusters show disciplinary and thematic affinities. Cluster 1 (red) contains journals primarily associated with work, gender, ethnicity, race and education. Cluster 2 (green) includes a strong urbanistic and geographic emphasis, suggesting that journals focused on spatial dimensions of segregation tend to cite each other more strongly than those associated with other disciplines. Cluster 3 (dark blue) contains a more diverse set of journals, including sociology and urban history. In turn, the betweenness centrality analysis (interactive network) shows what journals from different disciplines emerge as the most central nodes in segregation research, namely Urban Studies, followed by Urban Geography, Social Forces, Demography and Ethnic and Racial Studies (Table ST15 in SI).

Our analysis into the knowledge production networks in segregation research illustrate a richly interconnected field driven by thematic diversity, international collaboration, and disciplinary intersections. Through co-citation, co-authorship, and bibliographic coupling analyses, we observe how various disciplines contribute to a nuanced understanding of segregation, linking concepts across social, urban, and ethnic studies. This global collaboration network reveals prominent contributions from the US and UK, with significant clusters of activity across Europe, Australia, and beyond, reflecting a collective effort to explore the complex, multifaceted nature of segregation. Together, these networks underscore the expansive, multidisciplinary reach of segregation research and its reliance on collaborative knowledge exchange.

# An ontology of segregation

We have identified and mapped hundreds of forms of segregation across a diverse scientific literature encompassing 169 disciplinary fields, revealing the extraordinary connectivity between these forms. Given the complexity of this mosaic, how can we make it more comprehensible and valuable to the multidisciplinary community of researchers studying segregation and its many dimensions? Our approach has been identifying segregation forms and their relationships across over a century of literature. This search for a systemic understanding of segregation in its multiple manifestations is akin to an ontology. The term “ontology” originates from philosophy, where it refers to the study of existence. For example, Aristotle's ontology defines primitive categories like substance and quality, used to account for existing entities. In the early 1980s, Artificial Intelligence (AI) researchers adopted the term in computer and information science to describe both a theory of a modeled world and a component of knowledge systems (Guarino & Giaretta, 1995). An ontology defines a set of representational primitives, such as classes (or sets), attributes (or properties), and relationships, to model a domain of knowledge. This includes information about their meaning and the constraints on their logical application, similar to relational models for representing individuals, their attributes, and their relationships (Gruber, 2016).

Ontologies support cross-references and diverse domain-specific relationships such as depends-on, owns/belongs to, produces/is produced by, and has members/is a member of. As a networked structure, an ontology can model real-world domains where entities are connected through multiple types of relationships, properties, and constraints. This flexibility allows for richer semantic definitions, not just classifying entities but describing their relationships in greater detail and complexity (Hedden, 2016). Ontologies, therefore, provide deeper modeling of knowledge, capturing both the entities and the intricate connections between them.

In this paper, we propose an inductive approach to ontology creation. Additionally, we define the nature of the relationships and typological positions that various forms of segregation occupy within this conceptual space, based on the following definitions:

i. **Segregation form** refers to a specific process, practice or situation of separating or restricting interaction between individuals or social groups based on distinguishing characteristics, such as race, income, religion, or other social attributes. These forms are manifested through observable social, spatial, material, or economic patterns. Segregation forms are context-dependent, reflecting how segregation forces manifest within a particular environment, time, or population, shaping and being shaped by the surrounding societal, cultural, and economic conditions.

ii. **Segregation type** is a broader conceptual category or semantic group that encompasses multiple related forms of segregation. Types represent a generalization of shared underlying structures, processes or properties that may manifest through distinct but related forms. For example, residential segregation might be considered a type that encompasses various forms, such as income-based or ethnic-based residential segregation. The defining feature of a type is its ability to group specific forms based on common social, material or institutional mechanisms, allowing for general patterns of segregation to be identified across various contexts.

Segregation forms can intersect and belong to multiple types. For instance, "metropolitan Hispanic segregation" encompasses ethnic, geographic, urban and spatial segregation types (Fig. 17). The ontological method therefore avoids a strictly hierarchical structure making relationships exclusively vertical, as found in taxonomies in biology, opting instead for a richer relational approach. The method should also identify typological relationships between segregation forms and types from the bottom up, meaning that such relationships emerge from information produced or latent in the literature.

We employed a natural language processing (NLP) approach to group and rank SFs based on their semantic similarity using hierarchical clustering. First, the SFs were converted into high-dimensional numerical representations (embeddings) using a pre-trained sentence transformer model, specifically the all-mpnet-base-v2 from Sentence Transformers (Song et al., 2020). These embeddings capture the semantic relationships between the SFs. We evaluated multiple models, including SciBERT (allenai/scibert_scivocab_uncased) (Beltagy et al., 2019), BERT (bert-large-uncased) (Devlin et al., 2018), MPNet (sentence-transformers/ all-mpnet-base-v2) (Song et al., 2020), and T5 (t5-large) (Raffel et al., 2020), using combinations of distance metrics (cosine, euclidean) and clustering methods (ward, average, complete). The sentence-transformers/all-mpnet-base-v2 model is trained on large-scale datasets such as MultiNLI (for natural language inference), MS MARCO (for question answering and information retrieval), and TriviaQA (for question-answer pairs) (Hugging Face, 2021), enabling it to generate high-quality sentence embeddings by learning relationships between sentences across diverse tasks like semantic similarity, inference, and factual understanding.

After testing different configurations, we found that the MPNet model with cosine distance and complete linkage produced the most semantically meaningful clusters. This combination allowed for more distinct separations between groups, particularly in capturing the nuanced, multi-dimensional relationships between SFs. When computing semantic similarity for rare or specialised terms like "elderly residential segregation," general language models may struggle due to insufficient contextual understanding and poor representation of these terms in their training data. This can lead to inaccurate similarity scores, as the model may overemphasise more frequent components of the phrase (e.g. "residential" over "elderly") and fail to capture the nuanced meaning of the rare term. However, we found this only rarely to be the case in our application.

The complexity of semantic clustering for SFs lies in the fact that some SFs can theoretically belong to multiple clusters. For instance, ethnic residential segregation could cluster with both economic residential segregation (as they both address residential segregation) and ethnic school segregation (as they both involve ethnicity). Meanwhile, economic residential segregation and ethnic school segregation do not share a semantic commonality. This overlap in thematic relationships made it difficult to rely solely on traditional clustering metrics such as silhouette score or Davies–Bouldin index, which fail to account for such intersections. Manual evaluation, therefore, was necessary to assess the coherence and interpretability of the clusters. Multiple authors with expert knowledge qualitatively associated SFs with relevant labels, following predefined criteria developed in the coding phase of this research (see the Codebook) to reduce subjectivity and ensure consistent assessment. Clusters were assessed based on their ability to group SFs that shared similar meanings or contexts and assigned with labels. We identified 32 labels able to sufficiently represent such clusters of common features as segregation types (STs).

After clustering, we finalized the ontology. Each SF could belong to one or more types, with only one SF being assigned a maximum of eight types. Each ST is associated with a cluster as a node in its local network of directly related SFs. Since SFs can belong to multiple clusters, they form a network of relationships between the SFs and their types, culminating in an integrated ontology comprising 32 distinct segregation types. This method allows us to identify key groupings and map the overall relational structure of SFs. We produced a network graph (Fig. 17) using color and size to distinguish between SFs and types, making it easier to interpret how different forms of segregation correspond to specific ontological categories. This visualization highlights the differences in complexity among SFs, offering a clear view of their categorization and relationships.

The segregation ontology reveals a distribution where a few highly connected types dominate, while there is a gradual decline in frequency towards the tail end (Fig. 18a). Types like "social" (159 associated SFs), "racial" (127 SFs), and "spatial" (121 SFs) are most frequent, reflecting their widespread relevance across multiple forms of segregation in SR. In contrast, types such as "rural" and "environmental," with just four relationships each, suggest more niche applications that may warrant further exploration. This pattern indicates that certain types of segregation, like social and racial, permeate a broad range of contexts, while others, like environmental and rural, are tied to specific, less frequently studied scenarios. For high-frequency types, "social" encompasses forms such as social segregation, class segregation (also labelled economic), and religious friendship segregation (also labelled religious), reflecting broad social dynamics intersecting with economic, religious and a widespread of other dimensions. Similarly, "racial" includes racial segregation, statutory racial segregation (also labelled legal), and professional racial segregation (also categorized as occupational), demonstrating how racial divides are reinforced through legal and professional structures.

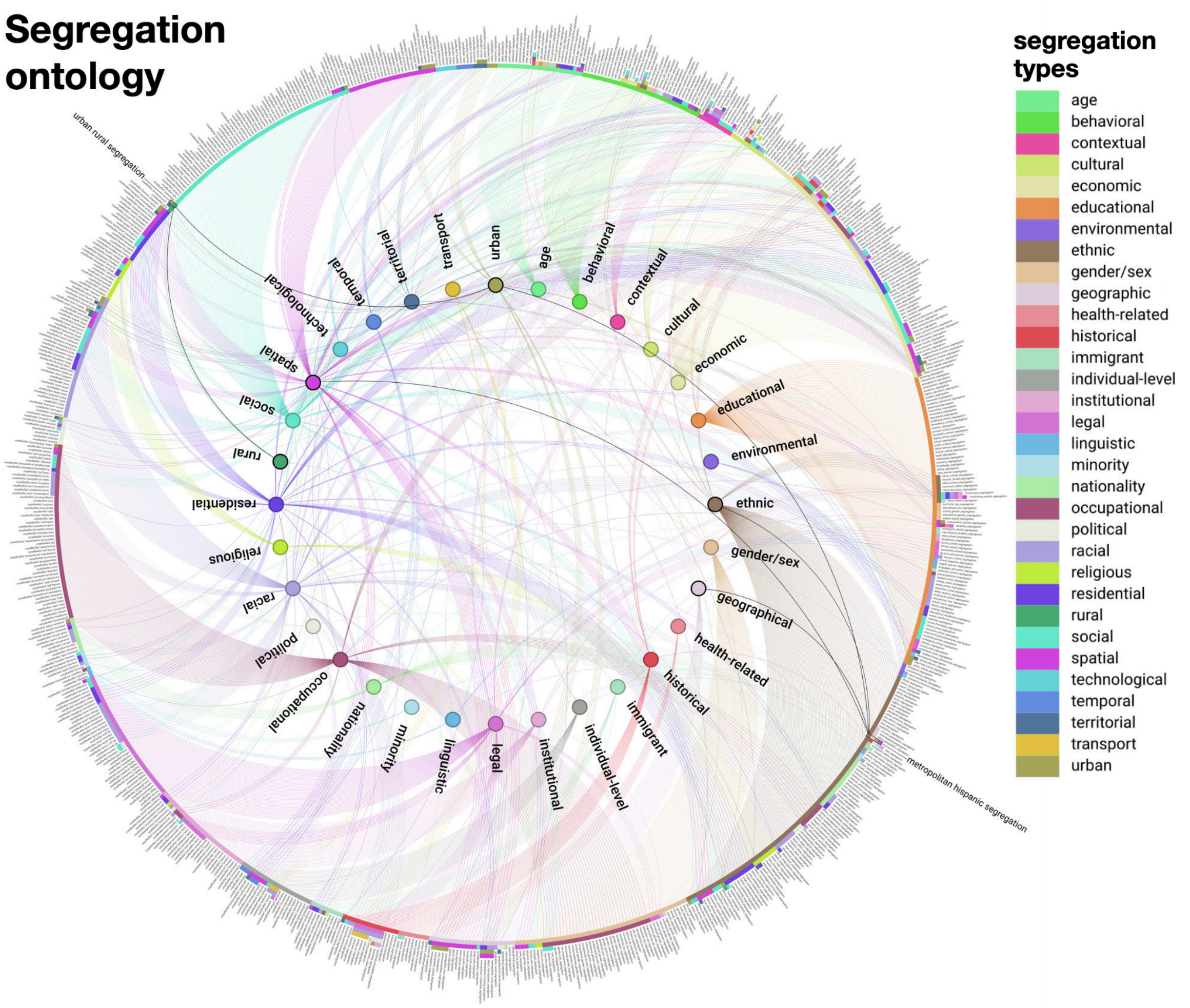

Fig. 17 – Network representation of the segregation ontology: the inner ring contains the 32 semantic groups or types, the outer ring contains the 804 identified SFs. Colors represent the types. The lines connect each type with all the SFs that they are associated with. The colored dots in the SF ring show the types associated with each SF. Two randomly selected SFs and their corresponding types are highlighted. View the zoomable high-resolution figure for more detail, or explore the interactive ontology, which is currently under development.

In contrast, the "rural" type includes SFs such as rural segregation, urban-rural segregation (also categorized under urban), and hukou segregation (also linked to ethnic and urban), which highlight specific geographical and societal divides. The "environmental" type covers forms like environmental segregation, ecological segregation, and racial ecological segregation (also categorized under racial), emphasizing the intersection of environmental and racial factors. These distinctions underscore how labels span a spectrum from broad societal divisions (social, racial) to more context-specific and potentially underexplored forms (rural, environmental). Overall, most SFs are associated with just one or two labels, indicating more narrowly defined concepts. However, exceptions exist, such as involuntary spatial segregation, which spans eight labels, including educational, ethnic, institutional, legal, and residential categories, reflecting the multifaceted nature of spatial inequalities. Similarly, postsoviet segregation and beach segregation span seven and five labels, respectively, showing their complexity. As illustrated in Figure 17, these highly-labeled forms are outliers, with the majority of SFs having fewer types, such as urban segregation and group segregation, which are each tagged with only one type, reflecting their more straightforward semantic categorization.

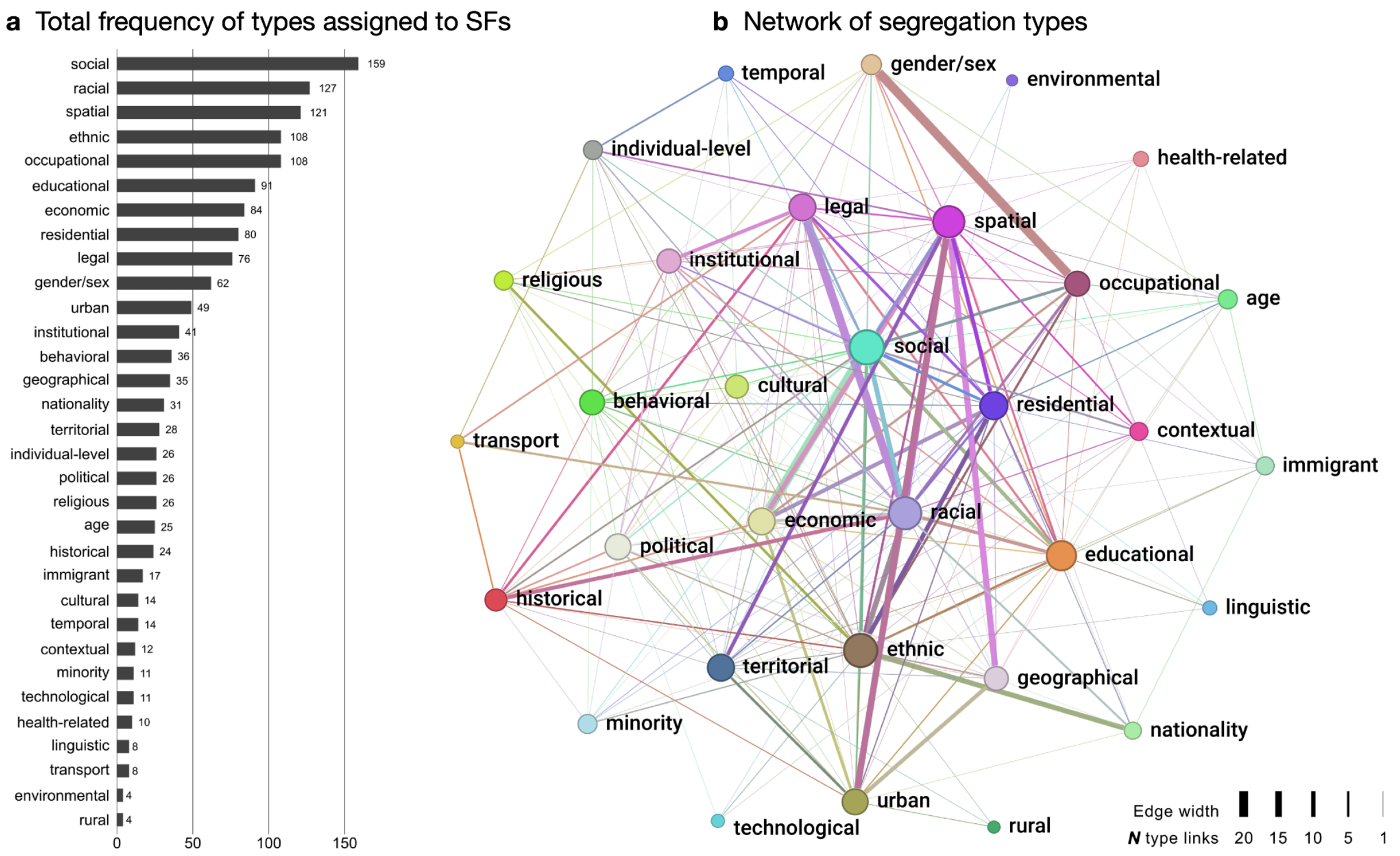

Fig. 18 – (a) Total frequency of types assigned to SFs.
(b) Semantic network of segregation types linked through cross-category SFs.

The total frequency of types assigned to SFs (Fig. 18a) reveals a highly skewed distribution, much like the earlier observations of individual SF frequencies. Within the network of segregation types (Fig. 18b), strong connections between certain types are dominant, highlighting frequent intersections in segregation forms. For example, the racial and legal types are linked 25 times, underscoring how legal frameworks often play a crucial role in enforcing or shaping racial segregation. Similarly, the link between social and economic types (25 semantic relationships) reflects the deep interdependence between social stratification and economic inequalities in driving segregation. The connection between urban and spatial types (23 relationships) points to the spatial and city-based dimensions that often underpin urban segregation dynamics. In contrast, more infrequent connections, such as urban and legal, political and temporal or religious and behavioral (each occurring just once), suggest that these types seldom overlap in the context of segregation forms. This highlights the specificity of certain intersections, which may reflect niche or rare applications in the study of segregation. Overall, the network demonstrates that, while certain types are strongly interconnected, the majority of type pairings are far less common, reflecting a similar skewed distribution in the broader network of segregation types.

This ontology of segregation (Fig. 17) facilitates a deeper understanding of the cross-disciplinary complexities and relationships within SR. It allows new work to be positioned within the context of existing forms and types, while also enabling the exploration of under-studied areas and promoting the development of new SFs and SR areas.

# Conclusions

Any attempt to develop a truly comprehensive transdisciplinary ontology of segregation must grapple with a fundamental complexity: segregation is inherently relational, manifesting through different mechanisms and material means. Its forms do not exist in isolation; they intersect and evolve through diverse social, economic, political, spatial and technological processes. Humans find ways to segregate. The forces driving segregation are persistent and adaptive, making them difficult to contain. Over the past 110 years, many facets of segregation have been explored across a wide array of disciplines. However, this diversity of disciplinary perspectives often includes conflicting definitions, conflations, and implicit assumptions**.** Yet, this same diversity offers ground for identifying overlooked or unexplored forms of segregation. Drawing from the Scopus database, we examined a vast body of segregation literature comprising 10,754 documents produced by 14,841 authors in 127 different countries, from 1913 to 2022 and covering 169 disciplinary fields. We conducted an extensive analysis of open-access sections of these documents to find key terms associated with "segregation", focusing on English-language bigrams and trigrams. Through this process, we traced the evolution of segregation research and uncovered patterns and intersections across disciplines:

I. *The exponential growth and diversification* of segregation forms is evident in the publication trajectories within segregation research (SR) over time (Figs. 1c, 3a-b, 5). We proposed mechanisms to account for this trend, theorizing that combinatorial and exploratory conceptual accretion (Fig. 4) plays a key role. Our analysis revealed that over 90% of trigrams created by combining terms from existing bigrams were published after their corresponding bigrams. This suggests that new segregation forms are often built on earlier conceptual foundations. The diversification rate surged dramatically in the 1950s, only to begin slowing in the late 1980s, indicating a gradual concentration of research around specific segregation forms. Indeed, the trend towards diversification coexists with the centrality of certain SFs in research. Notably, nearly 50% of the documents within top disciplinary fields in publication numbers focus on a few segregation forms (Fig. 10c).

II. *The structure and connectivity* within this highly heterogeneous field reveal underlying hierarchies and clusters of segregation forms, as shown by co-occurrence (CO) networks in the analyzed literature (Figs. 6-7). These networks highlight certain SFs' pivotal role in establishing thematic and empirical relationships across the field. Despite the inherent complexity of segregation and the diverse nature of research spanning nearly 170 disciplinary areas, we observed a surprising degree of connectivity between segregation forms. The many facets of segregation are remarkably related across the literature. This integration was unknown and is probably hard to be observed by researchers working on specific SFs. Journals from distinct disciplines are also closely linked through bibliographic coupling, underscoring the connections within the literature (Fig. 16b). For example, while many journals within the spatial disciplines form distinct clusters, our analysis revealed stronger-than-anticipated links between spatial and non-spatial approaches. This cross-disciplinary connectivity is further reinforced as segregation forms transcend disciplinary boundaries, contributing to the integrated structure of the field (Figs. 9-10).

III. *Shifting trends in multidisciplinarity, transdisciplinarity, and intersectionality* in segregation research reveal key transformations in the field. While segregation research remains inherently multidisciplinary, the number of disciplinary fields involved continues to grow (Fig. 8a). However, since the 1990s, both multidisciplinarity and transdisciplinarity indexes have been declining (Figs. 8c, 9a-b), as the publication of SFs has increasingly concentrated within specific fields (Fig. 8b). In contrast, intersectionality has shown a marked upward trend, particularly from the 2000s onward (Fig. 11b), reflecting the growing complexity of how different dimensions of segregation are being studied and understood in combination.

IV. *Contextual variations:* Our analysis revealed significant differences in the dominant forms of segregation across countries and regions, as local contexts naturally influence the research focus. For instance, racial segregation receives more intense scrutiny in North America and Africa, while income-related segregation is a prevalent focus in South America. In contrast, Europe, Asia, and Oceania tend to emphasize research on gender, ethnic, and occupational segregation, alongside other, more nuanced contextual differences (Fig. 12c). Although the United States has historically led segregation research, other countries and regions have steadily increased their contributions since the 1980s, reflecting a more globally distributed focus on the subject (Figs. 12a-b).

V. *Path dependence:* we hypothesized that historical trajectories of research – both in terms of time and geographic location – play a critical role in shaping the development of segregation studies. Our findings suggest that time, in particular, may influence the positioning of segregation forms within the co-occurrence topology (Fig. 13). This analysis also enabled us to trace the earliest connections between segregation forms (SFs) in the Scopus dataset (Fig. 14). The resulting co-occurrence network brings insights into how the identification of new SFs fosters connections between them. Our results indicate that while there is ample potential for conceptual innovation, path dependence remains a significant factor in the evolution of segregation research.

VI. *Knowledge production:* our analysis revealed the networks of co-cited works, highlighting the resulting clusters and the distribution of citations within the field (Fig. 15). We also examined the co-authorship network across countries, identifying the dominant role of the United States in knowledge production, alongside clusters of more closely collaborating countries (Fig. 16a). Additionally, we found that co-citing journals exhibit widespread connectivity, with clusters reflecting either thematic affinities or complementary research approaches (Fig. 16b).

Finally, we explored our findings to propose a bottom-up, network-based ontology of segregation, grounded in the relationships between identified forms – an effort that, to our knowledge, represents the first comprehensive attempt of this kind within the social sciences. The 32 identified segregation types serve as conceptual pathways, providing access to the vast array of segregation forms while revealing the semantic connections and intricate interrelationships that shape this complex landscape.

This meta-study of segregation involved the following aims and possibilities:

- *Clarifying the transdisciplinary field:* The proposed ontology provides a semantic framework to help organize and make sense of the vast heterogeneity of segregation forms, highlighting their relationships and potential for further conceptual development.
- *Supporting researchers across disciplines:* Our approach offers a structured overview of segregation forms, aiming to aid researchers from different fields in dealing with the relational nature of segregation and stimulating awareness of relevant explorations beyond disciplinary boundaries.
- *Promoting precision in terminology:* The empirical and conceptual mosaic rendered visible through this approach may foster a more deliberate use of terminology. It may encourage clearer conceptual formulations, reducing the risk of conflation between segregation forms and avoiding fragmentation or the unnecessary multiplication of terms that refer to the same segregation forms within and across disciplinary fields.
- *Encouraging exploration of the "adjacent possible":* Clarifying which segregation forms have been identified and uncovering their topology can stimulate researchers to explore new conceptual and empirical avenues. We hope these topological maps and the ontology will help researchers identify gaps in the literature, explore combinatorial possibilities, and extend their work into new, uncharted areas of segregation research.

The present approach carries certain limitations along with possibilities for future research:

- *Limitations of the Scopus database:* While Scopus provides extensive coverage of journals, book chapters, and conference proceedings, it does not encompass all academic publications. This limitation may affect the accuracy of our findings, particularly in relation to the year and country of first publication.
- *Inherent challenges of ontology-building:* The ontology we propose is likely to never be fully complete. This is not only because the field of segregation research will continue to evolve but also due to differences in how researchers interpret the textual units where forms of segregation appear.

- *Language limitations:* Our analysis generally did not capture segregation forms published in languages other than English. This may result in the exclusion of context-specific SFs and overlook the contributions of scholars working in non-English languages (see Acknowledgements). A critical next step for our research is to broaden the scope of works analyzed to include non-English publications, allowing for a more geographically and culturally diverse assessment of segregation forms and their origins outside the Scopus database.
- *Primary focus on segregation forms explicitly labeled as 'segregation':* our current approach does not account for related concepts (e.g. social isolation, spatial exclusion) or the implications of segregation (e.g. neighbourhood effects). However, we expect that most relevant forms not labeled as 'segregation' are captured within existing segregation n-grams, limiting the likelihood of substantive gaps in our bottom-up review and ontology.
- *Exclusion of fourgrams:* We omitted fourgrams from the analysis because the potential gains were deemed minimal compared to the interpretive effort required to accurately distinguish SFs from other four-word combinations ending in "segregation." Similar potential representations of segregation forms outside the "x segregation" and "x y segregation" structures were not captured. However, future methodological advancements may make it feasible to incorporate such representations.
- *Genealogies of segregation forms and their relationships:* The method we employed to map relations between SFs focused on how their connections unfolded over time through co-occurrences. We introduced such an approach to trace path-dependent trajectories, but it can be further refined to offer a deeper genealogical perspective on specific SFs and their conceptual evolution. We plan to explore these genealogies in upcoming qualitative work.
- *Expanded scientometric analysis:* Future research can extend this study by conducting more in-depth scientometric analyses of citation networks. This could further illuminate how SFs are discovered through combinatorial and explorative work, revealing new possibilities and helping identify emerging segregation forms.
- *Validation of the segregation ontology* requires review performed by domain experts that critically evaluate the ontology to ensure its accuracy, comprehensiveness, and alignment with existing segregation literature. This process includes iterative refinements based on expert feedback, emerging research trends, and additional data sources, to enhance the ontology's robustness and applicability.

To advance the study of the emergence and interconnections between segregation forms and concepts, we have launched the collaborative platform Segregation Wiki, featuring the 804 identified segregation forms. This resource is designed to be continually expanded and refined by the research community. Furthermore, we are actively working to make our ontology and data framework publicly accessible, enabling other researchers to explore, analyze, and further develop the complex landscape of segregation forms.

The ever-expanding recognition of facets of segregation is rooted in the nature of the phenomenon. But is there a limit to the discovery of new segregation forms? Our analysis suggests that, even though the number of new SFs identified every year continues to increase exponentially, the growth of entropy-based diversity of SFs has slowed down since the late 1980s, as new publications have increasingly focused on certain established SFs, while newly identified SFs have become more concentrated within specific disciplinary fields. While segregation is an intricate and evolving aspect of society, we must ask: will the discovery of new forms eventually diminish as theoretical frameworks increasingly diversify and align with the reality of segregation? Can segregation research fully capture the evolving landscape of segregation? At this stage, we cannot definitively answer these questions. As methods continue to evolve and the scope of analyzed literature expands, addressing these questions will be essential for future research.

**Acknowledgements**
We recognize the inherent biases in the global scientific production and dissemination landscape, which disproportionately disadvantage certain groups. Researchers who are native speakers of languages other than English, those working outside of wealthier countries or prestigious institutions, those with limited resources, women in science, non-white researchers, and scholars with disabilities, among others, are certainly underrepresented in the body of literature analyzed in this study.

Identifying the year of first publication for segregation forms using existing indexing systems brings significant methodological challenges, particularly because public databases do not fully capture the whole specialized literature, especially for works published in languages other than English. We empathize deeply with the authors whose valuable contributions may not have been included due to these limitations. If there are instances of innovative work that were overlooked, we sincerely apologize. We welcome any notifications of such cases and will make every effort to acknowledge them in the final version of this article and on the Segregation Wiki platform.

The authors would like to express their gratitude to Laura Vaughan, Sako Musterd, Kerli Müürisepp, Ilgi Toprak, Luisa Batista, Renato Emerson, and Vítor Oliveira for their valuable comments on earlier versions of this work. We extend special thanks to Renato Saboya for his significant contributions to conceptualization, coding, manuscript revision, and supplementary information, as well as to Camila Carvalho for her involvement in the early discussions on potential approaches. Any remaining issues are the sole responsibility of the authors. We thank CITTA UPorto and CEEC FCT (VN grant 2023.07510.CEECIND), FAPERJ (VN grant E-26/211.381/2021, CNPq (VN grant 315086/2020-3, MF grant E-26/201.573/2023) and CAPES (OP grant 88887.877125/2023-00) for financial support.

**Data availability**
Scopus data supporting this study's findings are available at https://www.elsevier.com/products/scopus.

**Code availability**
The code used for the analysis can be made available on request.

**Authors and Affiliations**
Vinicius M. Netto, Research Centre for Territory, Transports and Environment, University of Porto | CITTA FEUP, Portugal.
Kimon Krenz, The Bartlett School of Architecture, University College London | UCL, UK.
Maria Fiszon, Graduate Program in Architecture and Urbanism, Fluminense Federal University | UFF, Rio de Janeiro state, Brazil.
Otávio Peres, Graduate Program in Architecture and Urbanism, Federal University of Pelotas | UFPel, Brazil; Federal University of Santa Catarina | UFSC, Brazil.
Desirée Rosalino, National School of Statistical Sciences (ENCE), Rio de Janeiro, Brazil.

**Author contributions:** VN defined the research problem and contributed to conceptualization, methodology, investigation, writing – original draft, funding acquisition, project administration and supervision. KK contributed to conceptualization, methodology, software, formal analysis (NLP, network and ontology analysis), investigation, data curation, writing – review and editing, and visualization. MF contributed to conceptualization, literature review, methodology, investigation, and manuscript review. OP contributed to conceptualization, methodology, investigation, formal analysis (growth, diversity, multidisciplinarity, and intersectionality analyses), manuscript review and supplementary information. DR contributed to software, data curation, investigation, formal analysis (statistical and network analyses) and visualization.

**Competing interest declaration:** The authors declare no competing interests.
**Corresponding author:** vmnetto@fe.up.pt